\documentclass[preprint,12pt]{elsarticle}

\usepackage{pgfplots}
\pgfplotsset{compat=1.18}
\usepackage{amssymb}
\usepackage{amsmath}
\usepackage{float}
\usepackage{xspace}
\usepackage[bookmarks=false]{hyperref}
    \hypersetup{colorlinks,
      linkcolor=blue,
      citecolor=blue,
      urlcolor=blue}

\newcommand{\NBZ}{\mathcal{Z}\xspace}
\newcommand{\NbNoeudsCentraux}{\mathcal{N}\xspace}

\journal{Journal of Information Security and Applications}

\begin{document}

\begin{frontmatter}


\title{Concept drift mitigation through community and spectral graph analysis for the detection of cyberattacks in network traffic}

\author[1,2]{Julien Michel}
\ead{julien.michel@epita.fr}
\author[1]{Abdul Qadir Khan}
\ead{abdul-qadir.khan@epita.fr}
\author[1,2]{Majed Jaber}
\ead{jaberm@unistra.fr}
\author[1,2]{Pierre Parrend}
\ead{pierre.parrend@epita.fr}%% Author name

\affiliation[1]{organization={Laboratoire de Recherche de l’EPITA},%Department and Organization
            addressline={14-16 Rue Voltaire}, 
            city={Le Kremlin-Bicêtre},
            postcode={94270}, 
            country={France}}
            
\affiliation[2]{organization={ICube, UMR7357, Université de Strasbourg},%Department and Organization
            city={Strasbourg},
            postcode={67000}, 
            country={France}}

%% Abstract
\begin{abstract}
In network traffic, legitimate behaviours and attack techniques evolve jointly --- the phenomenon known as `concept drift'. Every detector is thereby left obsolete between two updates, and always one step behind adversaries. In this work, we propose to move the point of intervention from the model, repaired after the drift, to the feature space, selected before learning.
We therefore introduce \textit{t-robustness}, a stability score defined for each feature independently of any detection model, comparable across an entire feature space. It combines the step-by-step distance between successive statistical states of a feature, and its cumulative divergence from its initial state, so that a slow monotonic drift cannot pass for stability. The candidates are drawn from abnormal network connectivity patterns left by scans, DoS and communications between endpoints, read through graph community metrics and spectral metrics.
The evaluation is performed on the UGR16 dataset, across three learning scenarios and a control scenario, as well as without model update, and demonstrate that t-robust feature spaces sustain detection where the baselines collapse: retained expectancy at the last test interval reaches 0.6025, against 0.5230 for graph community features and 0.3831 for the base NetFlow features.
\end{abstract}

% \includegraphics{figures/michel.2026.jisa.graphical_abstract.png}

%  \item  Concept drift quantification at feature level — Introduces t-robustness, a model-independent score that measures the temporal stability of individual features.
%  \item Graph-based feature enrichment — Uses graph community and spectral metrics to capture attack-related changes in network connectivity and topology.
%  \item Proactive robust feature selection — Combines t-robustness with information gain to select features that are both stable over time and useful for cyberattack detection, avoiding continual model retraining.
%  \item Improved detection resilience — On the UGR16 dataset, t-robust feature spaces maintain detection performance under concept drift, reaching 0.6025 retained expectancy, compared with 0.5230 for graph-community features and 0.3831 for base NetFlow features.

%% Keywords

\end{frontmatter}



\section{Introduction}
\label{intro}

The detection of anomalies in evolving networks that are affected by concept drift \cite{gama2014survey} is a complex and open challenge \cite{komarchesqui2026comprehensive}. Internet access networks are a typical instance of such systems: user behaviours evolve continuously, and sustaining attack detection performance over time is a major challenge. We define \textbf{concept drift in detection} as the evolution of detection targets within an ever-changing environment over time \cite{webb2016characterizing}. This phenomenon is both a cause and a consequence of significant changes in data features. Concept drift is by nature tied to the feature space, that is, to the value space of communication packet parameters such as latency or size. The feature space can therefore be evaluated quantitatively over a given time frame. A feature space subject to high concept drift does not represent the general behaviour of the data; it only captures the patterns of the time period in which learning occurs. The challenge is therefore to identify a feature space which characterises the detection targets of the environment at any given time.

A feature space less vulnerable to concept drift supports the construction of more time-robust machine learning pipelines. The requirement for such a feature space is to represent attack behaviours in a way which remains distinct from the normal data of the environment over time: it maintains its discriminative power for attack identification independently of external factors. Existing approaches detect concept drift and then adapt the learning model, a process which requires frequent model updates and substantial time and resources \cite{agrahari2022concept}. The complementary question — which features remain discriminative over time — is largely left aside: reviews of concept drift in intrusion detection systems emphasise the lack of research addressing the intersection of concept drift and feature drift \cite{shyaa2024evolving}, although feature stability is a determining factor of the long-term effectiveness of machine learning models \cite{van2016test}.

We consider three families of candidates for time-robust features: original packet features, derived graph community metrics \cite{yang2012defining}, and derived spectral metrics \cite{jaber.25.softwarex}. Popular approaches for graph models for intrusion detection explore implicit feature analysis through embeddings \cite{bilot2023graph},through supervised or unsupervised learning models \cite{caville2022anomal}. However, recent evaluations demonstrate that the increase complexity of embeddings is not correlated with improved detection capabilities \cite{bilot2025sometimes}. Two motivations support the use of graph community metrics and spectral metrics. First, topology-based graph metrics have few dependencies in their construction, which makes them resilient to data evolution \cite{navruzov2022detection}: fewer dependencies entail greater robustness against concept drift. Secondly, attack behaviours affect network structures such as communities in a manner which differs from the heterogeneous activity of legitimate traffic. Graph-based metrics are therefore more resistant to attacker strategies, since they do not rely on information which can easily be reconstructed.

To define and evaluate robust feature spaces, we structure our work around three research questions:
\begin{description}
 \item [RQ1:] How can concept drift be quantified in communication data?;
 \item [RQ2:] How can this quantification be used to build a feature space which remains robust over time?;
 \item [RQ3:] Are graph community metrics and spectral metrics relevant candidates for building a long-lasting feature space for attack detection?
\end{description}

To answer these questions, we assess detection performance with feature sets enriched with graph community metrics and with spectral metrics in a concept drift context. We define metrics which quantify drift relatively to each feature of the feature space, and evaluate the stability of these features in the dataset. By combining graph community metrics and spectral metrics with feature stability metrics, we build a time-robust — hereafter t-robust — feature set, and evaluate its detection performance under concept drift. The proposal is validated over three learning scenarios and a control scenario on the UGR16 dataset \cite{macia2018ugr,UGR16}. Models relying on the t-robust set prove more stable over time and achieve a better retained expectancy than models relying on the base set, on graph community metrics or on spectral metrics alone: the proposed metric \textit{retained expectancy} has a value of 0.6025 at the last test interval of scenario 1, against 0.5230 for the graph community set and 0.3831 for the base set.

The remainder of this paper is structured as follows. Section~\ref{SoA} presents the related research. Section ~\ref{graphconnectivity} explains the impact of cyberattacks on network topologies and on graph metrics. Section~\ref{condrift} defines the metrics and the requirements of the proposed robust feature spaces. Section~\ref{metodo} provides the evaluation methodology and the implementation details. Section~\ref{eval} presents the results and the performance evaluation. Section ~\ref{discussion} discusses the contribution of the proposed feature spaces and elicits further research challenges. Section~\ref{conc} concludes the paper.

\section{State of the art}
\label{SoA}

Concept drift denotes a change in the relationship between the detection targets, \textit{i.e.} the elements of the data which carry the decision, and their environment, \textit{i.e.} the data which the detection model treats as irrelevant~\cite{webb2016characterizing}. It is to be distinguished from data drift, which is a shift in distribution between a training set and a test set~\cite{ali2024machine}: under data drift the underlying behaviours are preserved and only their frequency changes, whereas concept drift redefines the behaviours themselves. The consequence for detection is direct. A drift which leaves the observed distribution of the data unchanged can nonetheless invalidate the decision boundary, so that concept drift is at once harder to observe and more damaging than data drift. The literature answers this difficulty with a family of \textbf{Concept Drift Detection} (CDD) approaches. We review these approaches below, then examine how the security domain exploits them, before establishing the position from which our own contribution proceeds.

\subsection{Concept drift analysis}
\label{soa:analysis}

Concept drift detection methods differ by the observable they monitor: the distribution of the features, the parameters of the drift episode, or the geometry of the data cloud.
Three families localise, date and qualify the drift, yet none of them prevents it. Their common output is an event --- a drift has occurred --- whose only possible consequence is an intervention on the model.

The Parallel Histograms through Time (PHT) model~\cite{galmeanu2024concept} monitors the first of these. It follows, for each relevant feature, the evolution of its distribution within sliding windows centred on the mean, which localises drift both in time and across the feature set. Drift is thereby made visible feature by feature, but the model stops at this visualisation: the exploitation of the localisation it produces is left to the analyst.

QuadCDD~\cite{wang2024quadcdd} monitors the drift episode itself, which it describes with four parameters: the \textit{drift start}, the last point at which the current data profile still holds; the \textit{drift end}, the point at which the accuracy of the detection model falls below a chosen threshold; the \textit{drift type} --- incremental, abrupt, recurring or gradual --- which qualifies the transition between these two points; and the \textit{drift severity}, which quantifies through accuracy rates the impact of the episode on the model. These parameters serve to adjust the model to the episode which has just been characterised: stability in the data stream is obtained by reaction.

The Typicality and Eccentricity Data Analytics-based Concept Drift Detector (TEDA-CDD)~\cite{nunes2024concept} monitors the geometry of the data. It maintains a reference model, compares it with the current model by means of the Jaccard index and of Euclidean distances, and declares the current model abnormal --- hence a drift --- when the deviation becomes significant. Its value lies in dispensing with labels and in operating at low computational and memory cost, which makes it applicable to data streams. Its output, however, is an alarm bearing on the data as a whole and not a diagnosis at the level of the individual features. \cite{gupta2025generative} defines density-aware learning dataset selection assisted with unsupervised generative labelling to leverage data geometry for detection in the presence of concept drift.

Such an intervention is indeed required, since the efficiency of a detection system confronted with drift declines until detection performance drops sharply, both the detection target and its environment having moved~\cite{guerra2022android}. Mitigation therefore consists in modifying the way the model learns and adapts~\cite{lu2018learning}, and no adaptation strategy is universal: the appropriate choice depends on the model, on the type of drift and on its intensity. Performance-driven strategies, the most widespread, update the learning model when its accuracy degrades, and are consequently no more reliable than the detection model whose degradation they measure. Distribution-driven strategies instead measure the distance between two data profiles, with a distance chosen according to the type of drift. Both share one prerequisite: the moment of the drift and the process which generates it must be identified before any correction can be applied. Tailored solutions can be devised for a given drift scenario and a given learning model, but in real data streams this identification is computationally expensive and cannot be sustained over time. Robust, long-lasting and generalisable solutions are needed instead. Adaptive learning~\cite{yu2024online} is one attempt in this direction: it takes several data streams as input and uses the correlations between them as a stable parameter, which allows it to outperform baseline models on six of the eight datasets evaluated under concept drift. The gain is obtained by seeking an invariant rather than by tracking the drift --- a change of perspective which the present work carries over to the features.

The features are indeed the level at which an invariant may be sought, since they constitute the interface through which drift reaches the model. Feature spaces built over the same dataset do not undergo drift with the same severity~\cite{hinder2024feature}, which establishes the feature space as a design variable of the detection problem rather than as one of its givens. When the statistical evolution of the features becomes strong enough to degrade detection, \textit{feature drift} affects the subset of the feature space on which the learning model relies~\cite{barddal2017survey}. Feature drift is thus the measurable manifestation of concept drift, and the level at which a preventive treatment, rather than a reactive one, can be defined.

\subsection{Concept drift and stability}
\label{soa:stability}

The key property for concept drift analysis is stability. Kalousis et
al.~\cite{kalousis2007stability} measure the stability of the feature
preferences an algorithm expresses — weights, ranks, or a selected subset —
by similarity indices across perturbed training sets. Nogueira et
al.~\cite{nogueira2018stability} gave the question axiomatic form,
requiring a stability estimator to be fully defined,
strictly monotone in the variance of the selection, bounded independently
of the dimension, maximal only under deterministic selection, and corrected
for chance. Invariant risk minimisation \cite{arjovsky2019invariant} seeks,
one level higher, a representation whose optimal classifier holds across
given environments. All three perturb the sample or presuppose the
environments: none is indexed by time, and none returns a verdict on an
individual feature.

Hinder et al.~\cite{hinder2024feature} gives a formal notion of
concept drift grounded in a \emph{distribution process}. Let $T$ denote a
time domain and $\mathcal{X}$ the feature space of a data stream. A
distribution process is a pair $(D_t, P_T)$, where $P_T$ is a probability
measure over $T$ and $D_t$ is a family of distributions over $\mathcal{X}$
indexed by $t \in T$, such that each observation $X_i$ is drawn as
$X_i \sim D_{t_i}$ for its associated observation time $t_i \sim P_T$. The
process $D_t$ is said to exhibit \emph{drift} if there is a strictly
positive probability of drawing two time points with different underlying
distributions, i.e.
\begin{equation}
  P_{t_1,t_2 \sim P_T}\!\left[D_{t_1} \neq D_{t_2}\right] > 0.
  \label{eq:drift-def}
\end{equation}
Equivalently, drift holds if and only if the data $X$ and the observation
time $T$ are not statistically independent,
\begin{equation}
  X \not\perp\!\!\!\perp T,
  \label{eq:drift-indep}
\end{equation}
when $(X,T)$ is distributed according to the \emph{holistic distribution}
obtained by time-stamping each sample.

A \emph{drift detector} is then formalized as a decision procedure
\begin{equation}
  A_n : (T \times \mathcal{X})^n \to \{0,1\}
  \label{eq:drift-detector}
\end{equation}
operating on a finite sample, which is called \emph{valid} if the
asymptotic false-alarm rate on non-drifting streams stays strictly below
its asymptotic true-detection rate on drifting streams, and \emph{surely
drift-detecting} if it converges to the correct decision with probability
one as the sample size grows.

\subsection{Concept drift in attack detection}
\label{soa:attack}

Attack detection is a worst case for the reactive schemes described above, for three converging reasons. Attackers adopt new tools and new attack vectors as soon as these become available~\cite{liao2024multi}; they alter their tactics precisely because Security Operations Centres (SOCs) track them~\cite{sgandurra2016evolution}; and the environment is no more stable than the target, since new legitimate behaviours appear throughout the lifecycle of a network~\cite{adepu2016generalized}. Drift is therefore permanent, partly adversarial, and simultaneously affects the detection target and its background. A defence which waits for drift to be measured before reacting is by construction one step behind an adversary who chooses the moment of the change: the representativity of learning data is therefore a well identified challenge \cite{mwiga2026generative}.

Concept drift has received particular attention in malware detection, where classifiers trained on historical samples degrade rapidly as attackers evolve their techniques and the underlying data distribution shifts over time. Pendlebury et al. \cite{pendlebury2019tesseract} exposed how spatial and temporal biases in experimental design inflate reported performance, proposing the TESSERACT framework and the AUT metric to enforce temporally consistent evaluation and reveal the true, drift-degraded performance of malware classifiers. Building on this evaluative foundation, Barbero et al. \cite{barbero2022transcending} revisited the Transcend conformal-evaluation approach \cite{jordaney2017transcend}, showing that its original drift-detection thresholds were miscalibrated and introducing a more robust, statistically grounded rejection mechanism that better isolates drifting samples before they corrupt classifier decisions. Chen et al.  \cite{chen2023overkill} shifted the focus from the model to the feature space itself, analyzing which features are most responsible for drift in malware detectors and questioning whether the field's growing reliance on multi-source, feature-space explanations of drift is proportionate to the actual gains they deliver over simpler baselines.

A review of concept drift detection in intrusion detection systems (IDS)~\cite{shyaa2024evolving} ascribes the difficulty of detecting and categorising drift to three obstacles --- the high dimensionality of the data, class imbalance, and the optimisation of the learning model --- and designates the intersection of concept drift and feature drift as a blind spot of the field. The works which do reach the feature level confirm this diagnosis while remaining reactive in their principle. 
%The impact of feature drift on detection performance in data streams is examined in~\cite{chen2023overkill} on DroidEvolver~\cite{xu2019droidevolver}, an online learning model for malware detection: the feature set is adjusted from the analysis of the stream, and once drift is detected the features initially selected are re-evaluated and a new selection is performed with a linear SVM so as to follow the new data profile. 
The DI-NIDS (Domain-Invariant Network Intrusion Detection) framework leverages Domain-Adversarial Neural Networks (DANN) to tackle concept drift \cite{layeghy2023di}. After training on labelled and unlabelled date, the authors exploit the feature extraction branch of the DANN to obtain the domain invariant features. Then, they apply one-class Support Vector Machines or One-Class SVM (OSVM) on the extracted features to build a model for cross-domain anomaly detection.
The FeSAD framework~\cite{fernando2024fesad, fernando2024fesadphd} pursues the same objective for a learning-based ransomware detector, with the explicit aim of extending its lifespan: drift severity is quantified with the Heterogeneous Euclidean Overlap metric, the model is retrained on a core feature set, and a genetic algorithm generates a new set of features whenever drift intensity exceeds a predefined threshold~\cite{fernando2022fesa}. Recda framework \cite{yang2024recda} favours noise generation during the initial learning phase combined with manually crafted sampling at fine-tuning to reduce the dependency on labelled learning date.
In each of these cases, the feature space is repaired once the drift has been observed. Its stability is an outcome of the procedure; it is never a criterion applied before selection.

\subsection{Positioning of the contribution}
\label{soa:positioning}

The state of the art thus converges on a single loop: characterise the drift by its severity and its type, then update the model, at a cost in time and resources which recurs at each episode~\cite{agrahari2022concept}. The limitation of this loop lies not in the quality of its detectors but in its structure, which can only yield models that are already outdated at the moment they are replaced. Feature stability, whose role in the long-term effectiveness of learning models is established~\cite{van2016test}, offers a way out of the loop, on the condition that it be measured on the features themselves rather than inferred from the degradation of a model.

Two properties of the security context make such a measurement worthwhile. Concept drift is mediated by the feature space, and the resistance of features to drift is unequal within one and the same detection problem.
Attacks, for their part, are constrained by their objectives. An adversary
who alters the traffic her campaign emits is not free to alter it
arbitrarily: every modification must leave the campaign still capable of
achieving what it was mounted to achieve. This is the constraint which the
problem-space formulation of adversarial machine learning makes
precise~\cite{pierazzi2020intriguing}, and which the attacker models
proposed for network intrusion detection translate into the capabilities an
adversary realistically holds over network
traffic~\cite{apruzzese2022modeling}. Features which are simultaneously discriminative for attacks and stable over time should therefore be identified.

We consequently formulate the hypothesis which this paper sets out to demonstrate: \textit{a stable feature space can be constructed a priori, by quantifying the statistical evolution of the features --- their feature drift --- and by retaining those features which represent attack behaviours consistently over time}. Establishing this hypothesis requires three elements, which the remainder of the paper provides in turn: a family of candidate features whose construction depends little on the volatile parameters of traffic, namely graph community metrics and spectral metrics (Section~\ref{graphconnectivity}); a measure of drift defined at the level of the individual feature and independent of any detection model (Section~\ref{condrift}); and an experimental protocol which evaluates detection performance along the time axis rather than at a single point (Sections~\ref{metodo} and~\ref{eval}).

\section{Graph connectivity models}
\label{graphconnectivity}

Access networks are heterogeneous environments whose traces are generated by continuously evolving behaviours, among which attacks of several kinds occur. In an attack detection pipeline, the features are the first component over which the designer retains control: the representation chosen for the data determines what the learning model is in a position to discriminate. This section establishes the first of the three elements announced in Subsection~\ref{soa:positioning}, namely a family of candidate features whose construction rests on the topology of the exchanges rather than on the volatile parameters of the traffic. The argument proceeds in three steps. We first show that attacks inscribe in the connectivity of the network a small number of motifs which are dictated by their objective and which recur from one occurrence to the next (Subsection~\ref{gcm:topologies}). We then derive from these motifs two families of measurements: graph community metrics, which partition the graph and characterise the resulting groups (Subsection~\ref{gcm:communities}), and spectral metrics, which summarise the whole graph through the spectrum of its Laplacian (Subsection~\ref{gcm:spectral}). The demonstration is conducted on the UGR16 dataset~\cite{UGR16}.

\subsection{Impact of cyberattacks on network topologies}
\label{gcm:topologies}

Attacks observed in access networks differ in their operating methods, and our aim is to represent them in a way which exposes the process by which they proceed, so that the information essential to that process can be extracted from the representation itself. We therefore analyse two dynamic graph representations, which carry complementary information: in the first, nodes are machines, \textit{i.e.} IP addresses (\textit{IP/IP graph}); in the second, nodes are services, \textit{i.e.} pairs formed by an IP address and a service port (\textit{IP,Port/IP,Port graph}). The comparison between the two granularities is itself part of the analysis, since a motif invisible at one scale may be manifest at the other.

\begin{figure*}[!htbp]
    \begin{center}
    \includegraphics[width=\textwidth]{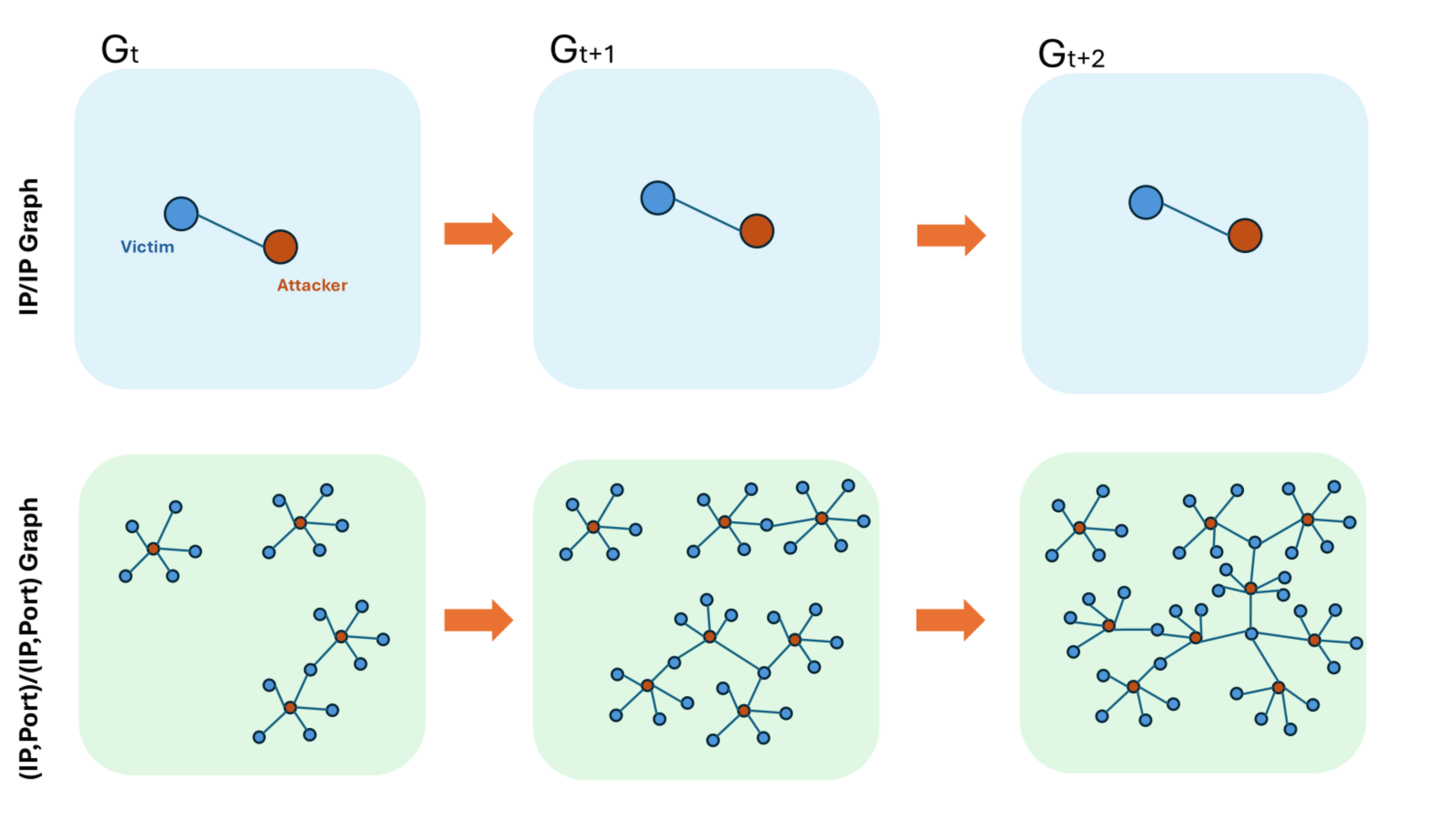}
    \end{center}
\caption{Graph representations of ports scans in UGR16 dataset (scan11 and scan44)}
    \label{fig:UGR_scan_graph}
\end{figure*}

Among the phases of the cyber kill chain~\cite{hutchins2011intelligence}, two govern what is observable in connectivity. Exploration, during which the attacker probes the target network, has reconnaissance as its purpose. `Actions on Objectives', during which the attack produces its actual effect, aims at altering one or several of the Confidentiality, Integrity and Availability (CIA) properties of the victim~\cite{samonas2014cia}: a breach of confidentiality is typically data theft, a breach of integrity the alteration of data or of configurations, and an attack on availability is generally a denial of service. Both phases require the attacker to establish exchanges which would not otherwise take place. We consequently restrict our attention to the attacks which create new connections and thereby modify connectivity patterns, since these are the attacks which graph connectivity models are able to expose: broadcast attacks such as scans and denials of service, and targeted attacks between otherwise unrelated nodes, such as endpoint-to-endpoint injections of malicious commands.

Reconnaissance collects information on a system in order to map it and to locate its vulnerabilities~\cite{bhuyan2011surveying}. Port scans belong to this category, their purpose being to identify the service ports left open on a device. Figure~\ref{fig:UGR_scan_graph} displays the dynamic graph representations of the port scans of the dataset. The \textit{IP/IP graph} associates one victim with each attacker and remains stable over the successive time steps: at this granularity the attack exhibits no discernible evolution. The same attack is unambiguous in the \textit{IP,Port/IP,Port graph}. Each attacker-victim pair is initially rendered by several components, most of them star graphs centred on the attacker; the remaining ones depart from the star by one or two edges while still being bipartite~\cite{asratian1998bipartite}. As the attack progresses, the number of attacker nodes increases, which reflects the opening of connections from successive ports of the same attacker address. The star components are correspondingly fewer and the bipartite components larger. The port scan is thus not merely visible at the service granularity: its progression is legible in the shift of the component population from stars towards larger bipartite structures.

\begin{figure*}[!htbp]
    \begin{center}
    \includegraphics[width=\textwidth]{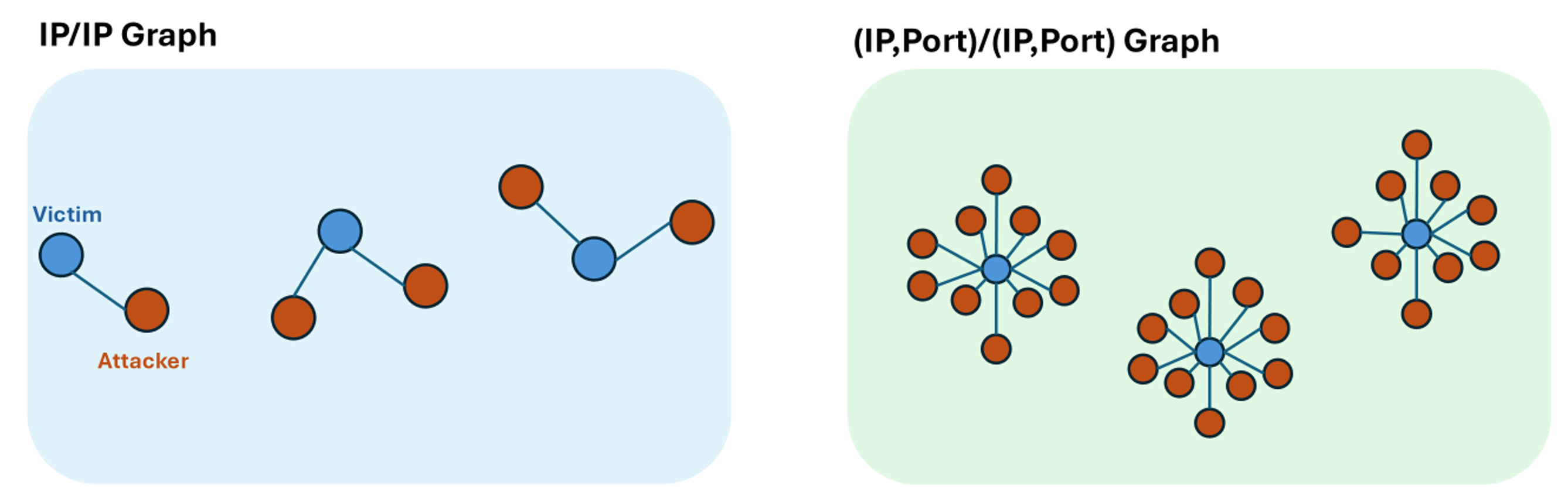}
    \end{center}
\caption{Graph representations of denial of service type of attacks in UGR16 dataset (DoS)}
    \label{fig:UGR_dos_graph}
\end{figure*}

Denials of service (DoS) constitute a second category. They are single-step attacks whose objective is to prevent a system or a service from operating~\cite{needham1993denial}, and which most commonly proceed by flooding the victim sub-network with connections issued from the attacker. The method is elementary, yet remains difficult to counter~\cite{sharma2023machine}. In the \textit{IP,Port/IP,Port graph} of Figure~\ref{fig:UGR_dos_graph}, the attack is characterised by an abrupt multiplication of attacker nodes converging on a single victim, that is, by a star structure whose central node is the target. The suddenness of the attack leaves no transition phase between the normal regime and the attack regime, and the star is already formed within a capture window of less than one minute. Where the scan is identified by the evolution of its motif, the denial of service is identified by the instantaneous appearance of one.

\begin{figure*}[!htbp]
    \begin{center}
    \includegraphics[width=\textwidth]{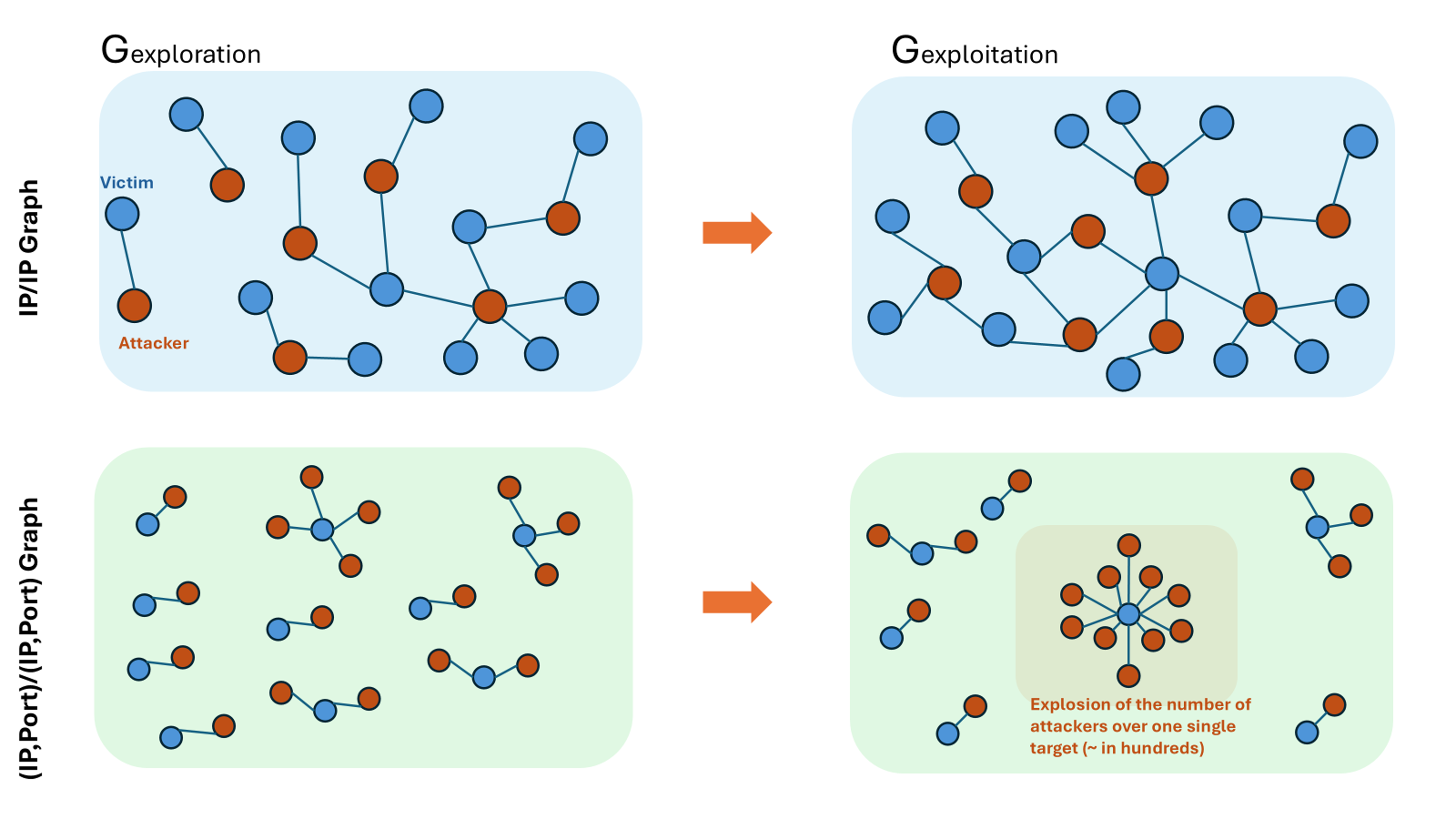}
    \end{center}
\caption{Graph representations of botnet type of attacks in UGR16 dataset (nerisbotnet)}
    \label{fig:UGR_botnet_graph}
\end{figure*}

A third category comprises the attacks whose behaviour differs from one time step to another. The UGR16 dataset contains such a multi-step attack in the form of botnet activity, the \textit{Nerisbotnet}, whose trace has been recorded and replayed in the dataset and whose dynamic graph is given in Figure~\ref{fig:UGR_botnet_graph}. Botnets are autonomous malware which traverse the whole kill chain: they identify new victims, perform an initial injection, establish persistence, then restart the cycle in order to propagate. Their behaviours are highly heterogeneous when different botnets are compared, but each botnet is highly regular in its own operation --- a dissociation which matters here, since regularity per family is what a stable feature can capture. All of them share the property of being installed on individual devices without the knowledge of their owners, the compromised devices then becoming attack vectors available to the operator. Two steps are distinguishable in both representations: an exploration phase and an exploitation phase.

During exploration, the \textit{IP/IP graph} exhibits one markedly large bipartite component, whose extremity nodes are the victims, together with a large number of smaller bipartite components; the corresponding \textit{IP,Port/IP,Port graph} only exhibits small star graphs centred on the victims.

During exploitation, the number of components involved in the attack decreases noticeably in the \textit{IP/IP graph}, to the point where a single bipartite component may account for the whole attack; as in the exploration phase, the extremity nodes of the largest connected component are almost always victims. In the \textit{IP,Port/IP,Port graph}, the components remain star graphs, with the exception of one singular star in which the number of attacker nodes surrounding the central victim node increases abruptly. In the dynamic graph, an exploitation phase is invariably preceded by several exploration phases, so that the succession of the motifs, and not only their presence, carries information about the attack.

Three properties emerge from these observations, and the remainder of the paper rests upon them. First, the motifs are few and recurrent: star components centred on an attacker or on a victim, bipartite components, and the fragmentation or coalescence of the component population account for everything observed here. Secondly, these motifs are imposed by the objective of the attack --- a scan must reach many ports, a denial of service must converge many sources on one target, a botnet must alternate propagation and exploitation --- so that an attacker cannot suppress them without abandoning the objective itself. Thirdly, their visibility depends on the granularity of the nodes: the port scan is invisible in the \textit{IP/IP graph} and manifest in the \textit{IP,Port/IP,Port graph}, whereas the botnet is legible in both, at different scales. The features derived in the following two subsections are consequently computed at both granularities, and are designed to measure the number, the size, the internal density and the temporal persistence of the components, rather than the value of any parameter of the traffic.

\subsection{Graph communities}
\label{gcm:communities}

Graph community metrics form our first family of candidates for robust features. Their construction has minimal dependencies, requiring only the end-to-end topology of the network and time~\cite{han2021evaluating}, and the community structures of a network topology are known to be tightly related to attack behaviours~\cite{rossetti2018community}. Both properties are those sought in Section~\ref{soa:positioning}: a construction insensitive to the volatile parameters of the traffic, and discriminative information specific to the attack itself.

Two types of graphs are generated from the flow data, here NetFlow captures, in accordance with the two granularities of Section~\ref{gcm:topologies}. In the first, nodes are IP addresses and each edge corresponds to an individual communication, that is, to one NetFlow record. In the second, nodes are pairs formed by an IP address and a port. The graphs are partitioned into communities, from which we compute the \textit{average degree} of the communities, their \textit{number of nodes}, their \textit{density}, their \textit{expansion}~\cite{yang2012defining} and the \textit{NEDIndex}~\cite{rahman2015nedindex}.

This choice of metrics follows the inventory of motifs established above. The number of nodes and the average degree measure the extent of a component and the concentration of its edges, hence the growth of the stars of a denial of service and the enlargement of the bipartite components of a scan. Density and expansion~\cite{yang2012defining} oppose an internally cohesive community to one whose edges point outwards, which is the formal counterpart of the distinction between a compact community of legitimate exchanges and the bipartite structure left by an attacker addressing many unrelated targets. The NEDIndex~\cite{rahman2015nedindex} completes the set to quantify the internal and external connectivity of communities.

The third property established in Section~\ref{gcm:topologies}, the temporal succession of the motifs, requires the metrics to be computed dynamically. We therefore apply time windows of 5 and 20 minutes and derive, for one and the same community between $t$ and $t+1$, the variations of the average degree ($\Delta degree$) and of the density ($\Delta density$). We further define \textit{Stability} as the ratio of similarity between two consecutive states of a community, $V_t$ denoting the set of nodes of the community at time $t$:
\begin{equation}
\label{eq: stability}
    Stability =\frac{|V_t\cap V_{t+1}|- |(V_t\cap\bar{V}_{t+1})\cup(V_{t+1}\cap\bar{V}_t)|}{|V_t\cup V_{t+1}|}
\end{equation}
Stability measures the persistence of a group of nodes across successive states, and thereby renders measurable the transitions which characterise multi-step attacks, such as the passage of the botnet from exploration to exploitation.

\subsection{Graph spectral analysis}
\label{gcm:spectral}

Spectral metrics form our second family of candidates \cite{martiny2026time}. Where community metrics describe the graph after it has been partitioned, spectral metrics characterise it globally, without any partitioning step, through the spectrum $\Lambda_t$ of its Laplacian at time $t$. The two families therefore provide complementary readings of the same connectivity, and their dependencies are of the same nature: the topology of the exchanges and time. The complete detection pipeline builds on spectral graph analysis~\cite{jaber2024graph}. Let $\lambda_{i}$ denote the $i^{\text{th}}$ Laplacian eigenvalue, $i \in [1,n]$, sorted in non-decreasing order, and let $\NBZ(t)$ be the multiplicity of the zero eigenvalue in $\Lambda_t$.

\par \textbf{Connectedness} quantifies the global interconnectivity of the network through its number of components, which the multiplicity of the zero eigenvalue provides directly. It thus addresses the fragmentation and coalescence of the component population observed for the scan and for the botnet. Connectedness is denoted $\mu_1(t)$ and is formulated as follows:
\begin{equation}
\label{eq: connectedness}
    \mu_1(t) = \frac{\exp{\frac{1}{\NBZ(t)}}}{\exp(1)}
\end{equation}

\par \textbf{Flooding} describes the behaviour of the low, non-zero part of the spectrum, which is associated with the most connected backbone of the graph, and therefore reflects the concentration of exchanges on the central nodes --- the signature of the star motif of a denial of service. Let $\NbNoeudsCentraux$ denote the number of central devices of the monitored network, such as switches and servers, and let $\mu_2(t)$ denote flooding:
\begin{equation}
\label{eq: flooding}
    \mu_2(t) = \left(\frac{1}{\NbNoeudsCentraux}\sum_{i=\NBZ(t)+1}^{\NBZ(t)+\NbNoeudsCentraux}{\lambda_i}^{(t)}\right) - 1
\end{equation}

\par \textbf{Wiriness} captures the variations of the upper tail of the spectrum, which reflect changes in dense or highly weighted interaction patterns. It is denoted $\mu_3(t)$:
\begin{equation}
\label{eq: wiriness}
    \mu_3(t) = \frac{1}{\NbNoeudsCentraux}\sum_{i=n-\NbNoeudsCentraux+1}^{n}\lambda_i^{(t)}
\end{equation}

\par \textbf{Asymmetry} measures spectral dispersion by counting the distinct eigenvalue gaps, and thereby tracks the structural evolution between two consecutive states of the graph, which is the spectral counterpart of the temporal succession of motifs. It is denoted $\mu_4(t)$:
\begin{equation}
\label{eq: asymmetry}
    \mu_4(t) = \mathit{Card}\{ i \geq 2 \; ; \; {\lambda_i}^{(t)} - {\lambda_{i-1}}^{(t)} > 10^{-12} \}
\end{equation}

These four metrics are computed on weighted graphs extracted from time windows built from raw traffic traces, the weights being derived from packet counts, byte volumes and rates. Each metric is thereby associated with an operational signal: Connectedness indicates the fragmentation or the merging of components, Flooding an abnormal load around the central devices, Wiriness a regime of high-intensity interaction, and Asymmetry a structural shift consistent with an attack-driven change of topology.

Community metrics and spectral metrics thus offer two complementary readings of one and the same connectivity, and both are computed from the topology of the exchanges and from time alone, without reading the value of any traffic parameter. This is what qualifies them as candidates for time-robustness. It does not, however, establish that they are time-robust: robustness is not a consequence of the construction of a feature but a property of its behaviour over time, which has to be measured. Defining that measurement is the object of the following section.

\section{Tackling concept drift}
\label{condrift}

Concept drift is well characterised in the literature, yet the metrics available to measure it do not answer the question raised in Section~\ref{soa:positioning}. They report drift either through the degradation of a detection model, which makes the measurement dependent on the very model whose obsolescence is at stake, or through a distance between two profiles of the whole dataset, which returns a single verdict for the entire feature space. Neither form allows features to be compared with one another, which is precisely what the construction of a robust feature space requires. This section supplies the missing instrument: a quantification of drift defined feature by feature and independent of any detection model, together with the conditions under which such a quantification can be evaluated. Four metrics are introduced for this purpose --- \textit{median-centered cut}, \textit{state distance}, \textit{t-equivalency} and \textit{t-robustness}. They form a chain in which each level aggregates the preceding one, so as to provide a single bounded score per feature.

\paragraph{Quantifying concept drift}

The quantification proceeds in two steps: the feature states of the dataset under analysis are extracted, then stability metrics are computed over these states in order to quantify the drift itself.

\textbf{Feature states} represent the statistical state of a feature over a time interval $\delta_t$. Two states are retained. The standard deviation $\sigma$ summarises the dispersion of the values. The \textit{median-centered cut} summarises their distribution: it is a clustering of the values taken by the feature during $\delta_t$, parametrised by the number of clusters $n$ and by the percentage difference $p$ from the median, a larger $n$ yielding a finer cut. The cut is centred on the median rather than on the mean, which keeps the state representative of the ordinary regime of the feature in traces where extreme values are frequent.

\textbf{State distance} is the proportional distance between two feature states, and the first of our stability metrics. It compares the two states bin by bin and sums the absolute differences of the bin proportions, so that a distribution which has been displaced and one which has been reshaped are both detected. With $pt1_k$ the proportion of bin $k$ at time $t1$, $pt2_k$ the proportion of bin $k$ at time $t2$, $n$ the number of bins and $k$ the index of the bin, whose content depends on the proportional distance from the median:

\begin{equation}
\label{eq:state_distance}
Sd= \sum^n_{k=1}{|pt2_k-pt1_k|}
\end{equation}

Computed between two consecutive intervals, \textit{state distance} measures a step, and a step alone. A feature may vary little from one interval to the next and nonetheless move far from its original behaviour by accumulation, which no local measurement detects.

\textbf{t-equivalency}, or time-equivalency, closes this gap. It is the average proportional difference of the \textit{median-centered cut} over $n$ intervals $\delta_t$, taken with respect to the first state, and is likewise a stability metric. With $Sd_k$, $1 < k \leq n$, the \textit{state distance} between state $k$ and state $1$:

\begin{equation}
\label{eq:t-equivalency}
equiT= \overline{|Sd_k|}
\end{equation}

Where \textit{state distance} measures the step, \textit{t-equivalency} measures the distance covered since the origin. The two are complementary and neither subsumes the other: an abrupt drift followed by a new stable regime is visible to the first and attenuated in the second, whereas a slow monotonic drift is invisible to the first and accumulated by the second.

\textbf{t-robustness}, or time-robustness, converts these measurements into the selection criterion the methodology requires. Three constraints govern its definition: the score must be high when the feature is stable rather than when it drifts, it must be comparable from one feature to the next, and it must not be granted to a feature which is stable in one respect only. It is defined as the minimum between, on the one hand, the average \textit{state distance} averaged with the average standard deviation difference and, on the other hand, the average \textit{t-equivalency}, over a distribution reduced to the $[0,1]$ interval. With $1 \leq k < n$, $Sd_k$ the \textit{state distance} and $Stdd_k$ the standard deviation difference, reduced to the $[0,1]$ interval, between state $k$ and state $k+1$, and $equiT_{k'}$ the \textit{t-equivalency} reduced to an interval of $t$ time windows from state $k'$ with $1 \leq k' < (n - t)$:
\begin{equation}
\label{eq:t-robustness}
t-robustness = \min(\frac{1-|\overline{Stdd_k}|}{2}+\frac{1-|\overline{Sd_k}|}{2},1-{equiT_{k'}})
\end{equation}

\texttt{t-robustness} estimates the extent to which the \emph{marginal} distribution $D_t(X_j)$, pooled over both attack and benign classes, remains invariant across the observation windows used in our pipeline.
It is a \emph{virtual-drift} measure: it quantifies
feature-wise stability of $P(X_j)$ over time, independent of any label information or downstream model. This places our metric in the same family as the feature-wise, marginal two-sample tests surveyed by Hinder
et al.\ (e.g.\ window-wise Kolmogorov--Smirnov statistics) \cite{hinder2024feature}, rather than in
the family of methods that test the conditional dependence $Y \not\perp\!\!\!\perp T \mid X$.
Consequently, a feature selected as $t$-robust is guaranteed to
have a stable marginal distribution, which makes no statement about the stability of relationship to the
attack label.
It is consequently possible, in principle, for a
$t$-robust feature to carry real drift --- a shift in
$P(Y \mid X_j)$ that leaves $P(X_j)$ unchanged.
We therefore adopt the working hypothesis, consistent
with prior feature-drift literature~\cite{barddal2017survey}, that
feature-level marginal stability is a computationally tractable proxy for the robustness of the induced classification boundary.

Two properties of this definition matter for what follows. The minimum is retained rather than the average, so that a feature is declared robust only if it is stable step by step, in dispersion as in distribution, \textit{and} stable with respect to its initial state: a feature which drifts slowly but steadily is penalised by the second term even though the first would absolve it. The score is bounded in $[0,1]$, which makes features rankable against one another and admits a threshold, and is the property on which the selection procedure of Section~\ref{metodo} rests.

\paragraph{Requirements to evaluation of robustness to concept drift}

A quantification is only as conclusive as the protocol which exercises it. We therefore state the requirements which an evaluation of robustness to concept drift must satisfy, and the conditions which the dataset must meet for the evaluation to carry any weight.

Three requirements bear on the evaluation. \textit{Quantifying feature drift}: a stability measure must be available for every feature of the dataset, so that features can be compared and not merely flagged. \textit{Performance of the feature space through concept drift}: the behaviour of attack detection models using a given feature space must be observed while drift occurs, and not only before and after it. \textit{Feature quality relative to concept drift}: for each feature, within a given detection environment, it must be decidable whether the feature belongs to the robust feature set. This third requirement does not reduce to the first. Stability is necessary but not sufficient: a feature whose value never changes is perfectly stable and contributes nothing to detection. Selection must therefore weigh the stability of a feature against its usefulness for the detection objective, which is what the methodology of Section~\ref{metodo} implements.

Three conditions bear on the dataset. \textit{Continuity}: the data must cover a sufficiently long period and contain enough records within it, the operational criterion being that the evaluation of a model displays decreasing performance over time. \textit{Time distance}: the start and the end of the period must be far enough apart for significant changes of behaviour to appear. \textit{Feature drift}: the original feature space must itself exhibit some degree of drift, if only partial, since a feature space which does not drift cannot distinguish a robust selection from an arbitrary one.

These requirements and these conditions together define what a reproducible evaluation of concept drift mitigation measures demands. The protocol and the dataset which satisfy them are presented in the following section.

\section{Methodology and implementation}
\label{metodo}

\subsection{Methodology}
\label{metodo:methodology}

\begin{figure*}[!htb]
\centering
\includegraphics[width=\textwidth]{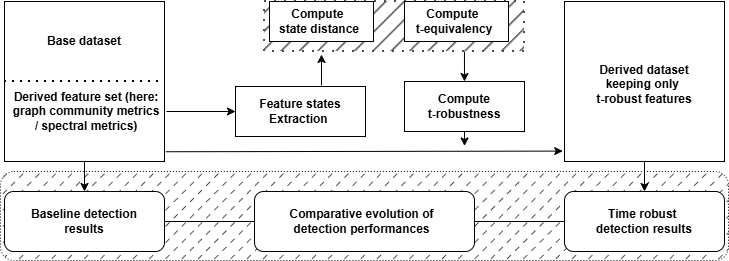}
\caption{Methodology and interaction between the models and the data}
\label{fig:method_schema}
\end{figure*}

The approaches reviewed in Section~\ref{SoA} identify the type and the severity of a drift and then retrain the learning model, at a recurring cost in time and computation. Our methodology replaces this loop by a selection performed once, upstream of learning, and must therefore establish two things at once: that a feature space selected on stability criteria degrades less over time than the space from which it is drawn, and that this gain is not obtained at the price of the detection performance itself. The protocol described below is organised around this double requirement, and is illustrated in Figure~\ref{fig:method_schema}.

The pipeline comprises four stages. The base dataset is first enriched with the derived features which Section~\ref{graphconnectivity} designates as candidates, namely graph community metrics, extracted here with the Louvain algorithm~\cite{louvain}, and spectral metrics, derived through the Laplacian Matrix of the graph. Feature states are then extracted from the enriched dataset at regular time intervals, and the drift of each feature is quantified by the \textit{state distance} and \textit{t-equivalency} measures of Section~\ref{condrift} between consecutive intervals. The \textit{t-robustness} of each feature is computed from these two measures, and features are ranked accordingly. Features whose \textit{t-robustness} falls below a predefined threshold are finally discarded, which yields a reduced, time-robust feature set.

Ranking on \textit{t-robustness} alone would satisfy the stability requirement and fail the performance one, for the reason stated in Section~\ref{condrift}: a feature which never varies is maximally stable and carries no information. Selection therefore combines two criteria --- the magnitude of the drift of a feature, measured by its \textit{t-robustness}, and its relevance to detection in the base scenario, measured by information gain. Only the features which satisfy both are retained as candidates for models whose performance is to be sustained over time.

The experimental design isolates concept drift as the explanatory variable. Several scenarios are constructed, each subjecting detection models to a different degree of concept drift; within a scenario, learning and evaluation conditions are held identical and the learning model is held fixed, so that the feature space is the only object which varies between the models compared. Two readings follow from this design. Read across feature spaces at constant scenario, the comparison identifies the features which contribute to temporal robustness. Read across scenarios at constant feature space, it validates the presence of drift in the data --- a performance-driven verification which does not presuppose the distribution-driven measurement, and which is therefore independent of it. The selected features are then used to retrain models in all scenarios, and their behaviour is compared with that of the models built on the base feature set.

The purpose of the methodology is thus to dispense with the detection of drift in real time and with the continual retraining of models in streaming environments. Features which are inherently more robust to temporal drift within given attack scenarios allow detection models to maintain a reliable performance over extended periods.

The effectiveness of the approach is assessed with three quantities, all built on the Matthews Correlation Coefficient (MCC), which we retain because it accommodates the class imbalance of the dataset~\cite{matthews1975comparison}, which have less than 2\% of attack data. \textit{Worst-case performance}: for each detection model and each learning scenario, the MCC is computed over every time period and the worst case is reported, which characterises the floor of the model rather than its average. \textit{Retained expectancy}: for each model, the ratio of the performance at each time interval to the initial performance is tracked, which expresses the capacity of the model to retain its detection capability over time. \textit{Detection rate difference}: for each prediction, the difference with the initial detection rate is computed across the intervals.

The last two quantities are given a definite form. The \textbf{MCC rate difference}, defined by analogy with the accuracy rate difference~\cite{lu2018learning}, measures learning stability in absolute terms:

\begin{equation}
\label{eq:MCC_rate_difference}
\Delta MCC= {\frac{MCC_1-MCC_2}{MCC_1}}
\end{equation}
The \textbf{retained expectancy rate difference} measures it in relative terms, over the whole sequence of intervals rather than between two of them:
\begin{equation}
\label{eq:retained_expectancy_rate_difference}
\Delta Retained\ expectancy= { 1-\frac{ \frac{\sum_1^k{MCC_k}}{k}}{MCC_1}}
\end{equation}

Both quantify the degradation of detection performance under concept drift, and it is on this degradation, rather than on performance measured at a single point, that the long-term robustness of a feature set is judged.

\subsection{Implementation}
\label{metodo:implementation}

The framework is evaluated over three learning scenarios and one control scenario (3+1). The three learning scenarios differ in the reference period used for the initial training, after which the detection algorithms are exposed to concept drift without any model update; they thus differ in how much of the past is known to the model when the drift begins. The control scenario removes the drift altogether: training and testing are performed over the whole time span of the dataset, which is partitioned into training and test sets. The two families of scenarios play distinct roles in the demonstration. The learning scenarios measure the resistance of a feature space to drift; the control scenario measures the performance that the same feature space attains in the absence of drift, and therefore establishes whether the robustness observed elsewhere has been paid for by a loss of discriminative power.

\paragraph{Analysis scenarios}
\label{subsec:scenarios}

Detection performance is evaluated with an XGBoost model~\cite{chen2016xgboost}, applied in turn to the base dataset and to the enriched feature sets containing either graph community metrics or spectral metrics.
\par{\textbf{Learning Scenario 1:}} the first two days of the UGR16 dataset are assumed to be known. The model is trained on these data, then applied to the remainder of the dataset without update.
\par{\textbf{Learning Scenario 2:}} the first five days, corresponding to the fifth week of July in the dataset, are used for training. The model is then applied to the rest of the data as in Scenario 1. Scenarios 1 and 2 differ only in the volume of the initial training period, which isolates the contribution of that volume to temporal robustness.
\par{\textbf{Learning Scenario 3:}} the model is tested on each segment of the UGR16 test dataset, the training data being always drawn from the immediately preceding time frame. This scenario assumes full knowledge of the preceding periods and therefore reproduces the continual retraining strategy of the state of the art.
\par{\textbf{Control Learning Scenario:}} this scenario represents a setting free from concept drift. Each data segment, except the first, is split into 80\% training and 20\% test sets, with 5-fold cross-validation to consolidate the results.

The 3+1 scenarios are exercised twice: first on the base and enriched feature sets, in order to establish the baseline behaviour and to verify that the data do drift, then on the time-robust feature set obtained by the selection procedure of Section~\ref{metodo:methodology}, the control scenario being reported as Scenario 4. Comparing the two passes yields the evolution of detection performance across scenarios and quantifies the gain attributable to the robustness-driven selection of features.

\paragraph{Concept drift in UGR16 dataset}
\label{sub:c_drift_ugr16}

The UGR16 dataset~\cite{UGR16} is used throughout the evaluation. It comprises network traffic collected between the fifth week of July 2016 and the fourth week of August 2016, and satisfies the three conditions stated in Section~\ref{condrift}.

It satisfies continuity, since it offers an uninterrupted time series over its period, with no missing interval, which allows detection performance to be compared across time segments; a dataset lacking such continuity, or covering too short a span, cannot expose the effects of concept drift at all. It satisfies time distance, its span of two months being sufficient for behaviours to change appreciably between the two ends of the period. It satisfies feature drift, since its background traffic is drawn from real network traces collected by an Internet security provider: the evolution of the environment is that of an operational network and not an artefact of the construction of the dataset, which is what makes the observed drift representative of the drift encountered in practice.

For the purposes of the experiments, each weekly file of the UGR16 test dataset is split into two segments along the time axis. Every segment except the first in chronological order is used as a test frame in the evaluation.

\section{Evaluation}
\label{eval}

The evaluation establishes, in four steps, whether a feature space selected on \textit{t-robustness} criteria sustains the detection of cyberattacks in network traffic under concept drift, in a classification task marked by pronounced class imbalance.
The four steps answer four distinct questions. The \textit{baseline performance assessment} determines what each feature space achieves when no drift occurs, by evaluating models trained under the control scenario. The \textit{robust feature selection} applies the methodology of Section~\ref{metodo} to identify the features which are temporally robust. The \textit{robustness evaluation under concept drift} measures the behaviour of the resulting feature space over the three learning scenarios of Section~\ref{subsec:scenarios}, the control scenario being retained as a reference. The \textit{analysis of feature space properties} finally computes the statistical properties of the robust feature spaces obtained, so as to independently corroborate the performance measurements.

\subsection{Baseline}

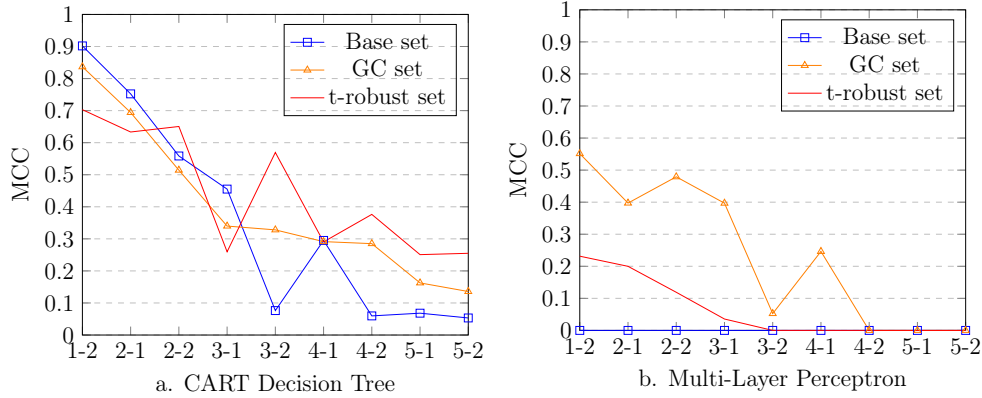
\begin{figure*}[!htb]
\begin{tabular}{cc}
\multicolumn{2}{c}{ 
\resizebox{0.47\linewidth}{!}{\begin{tikzpicture}
\begin{axis}[xlabel={a. CART Decision Tree},
ylabel={MCC},
xmin=0,
xmax=8,
ymin=0,
ymax=1,
xtick={0,1,2,3,4,5,6,7,8},
xticklabels={1-2,2-1,2-2,3-1,3-2,4-1,4-2,5-1,5-2},
ytick={0,0.1, 0.2, 0.3, 0.4, 0.5,0.6,0.7,0.8,0.9, 1},
legend pos=north east,
legend style={fill=none},
ymajorgrids=true,
grid style=dashed]
\addplot[
,color=blue,
mark=square,
]
coordinates {
(0,0.9018820058258056)
(1,0.7522578882764401)
(2,0.5584840912687199)
(3,0.4552004955356762)
(4,0.07628040437517049)
(5,0.29508959716944955)
(6,0.05965346689613711)
(7,0.06812792735708154)
(8,0.05322696361481105)
};
\addlegendentry{Base set}
\addplot[
,color=orange,
mark=triangle,
]
coordinates {
(0,0.8367044705811223)
(1,0.694091779398161)
(2,0.5142178407094489)
(3,0.33986683249783584)
(4,0.3280132173785804)
(5,0.29136806902556805)
(6,0.2850588061731295)
(7,0.16259476858160155)
(8,0.1357206915285324)
};%0.47774143521
\addlegendentry{GC set}

\addplot[
,color=red,
mark=cross,
]
coordinates {
(0,0.7029561004962579)
(1,0.6332867743009889)
(2,0.6505175675268581)
(3,0.25966163890057603)
(4,0.5694953115616586)
(5,0.29047631095417203)
(6,0.37625602250942514)
(7,0.2507701619810657)
(8,0.2549093819737676)

};
\addlegendentry{t-robust set}

\end{axis}
\end{tikzpicture}}
\resizebox{0.47\linewidth}{!}{\begin{tikzpicture}
\begin{axis}[xlabel={b. Multi-Layer Perceptron},
ylabel={MCC},
xmin=0,
xmax=8,
ymin=0,
ymax=1,
xtick={0,1,2,3,4,5,6,7,8},
xticklabels={1-2,2-1,2-2,3-1,3-2,4-1,4-2,5-1,5-2},
ytick={0,0.1, 0.2, 0.3, 0.4, 0.5,0.6,0.7,0.8,0.9, 1},
legend pos=north east,
legend style={fill=none},
ymajorgrids=true,
grid style=dashed]
\addplot[
,color=blue,
mark=square,
]
coordinates {
(0,0)
(1,0)
(2,0)
(3,0)
(4,0)
(5,0)
(6,0)
(7,0)
(8,0)

};
\addlegendentry{Base set}
\addplot[
,color=orange,
mark=triangle,
]
coordinates {
(0,0.5514264493369874)
(1,0.3971695264571866)
(2,0.4791263555327967)
(3,0.39684275989725015)
(4,0.05252678333065942)
(5,0.2462995795958863)
(6,0)
(7,0)
(8,0)
};%0.25502848612
\addlegendentry{GC set}

\addplot[
,color=red,
mark=cross,
]
coordinates {
(0,0.23151926184423796)
(1,0.19990322665961074)
(2,0.11896966695366931)
(3,0.035531454863701475)
(4,0)
(5,0)
(6,0)
(7,0)
(8,0)

};
\addlegendentry{t-robust set}

\end{axis}
\end{tikzpicture}}}
\end{tabular}
\caption{MCC evolution over learning scenario 1 for base, graph community, and t-robust sets with CART and MLP models. The x-axis denotes the capture period (week-segment notation N-M) according to scenarios with N the capture week and $M \in {\{1, 2\}}$ the week period.}
\label{fig:scenario1_CART_MLP}
\end{figure*}

The results reported in the remainder of this section are obtained with XGBoost. Before exploiting them, we verify that the phenomena observed are not an artefact of that choice of learning model. Figure~\ref{fig:scenario1_CART_MLP} gives the evolution of the MCC of two further models, CART~\cite{rutkowski2014cart} and MLP~\cite{taud2017multilayer}, over the same learning scenario 1 and the same feature sets as the XGBoost model of Figure~\ref{fig:mcc}.a. The CART model behaves in the same way as the XGBoost model, with slightly lower initial performances and a marked drop at 3-1, where the t-robust model falls to an MCC of $0.2596$; it nonetheless achieves the best MCC of all models of scenario 1 over the 3-1, 4-2, 5-1 and 5-2 periods. The MLP model performs poorly throughout, which reflects the high specificity of neural network based models rather than a property of the feature spaces being compared. Two further models could not be evaluated: KNN~\cite{steinbach2009knn} could not be fitted on the graph features within a reasonable time given our computational resources, and E-GraphSAGE~\cite{lo2022graphsage}, although considered, proved unsuited to the scale of our graphs, having been designed for a high number of edges over a few dozen nodes whereas the graphs considered here are sparse. The comparison therefore supports the transposition of our results to CART, the remaining families of models being out of reach under our constraints.

\subsection{Graph structural metrics}
\label{eval:structuralmetrics}

We evaluate our models, which we build over different feature sets, on four different scenarios. The \textit{Base set} is comprised of the $Packets$, $Bytes$, $Duration$, $Destination Port$, $Source Port$ and $ToS$ features. The \textit{Base set} evaluations are represented by the blue curves of Figure~\ref{fig:mcc}. Likewise, the green curves represent the \textit{Graph set} which include the features of the \textit{Base set}, enriched with two graph structural metrics: the $Average\_node\_degree$ and the $Density$ of the whole graph. This two metrics translate to eight new
features over, the two type of graphs and the two time windows of 5 and 20 minutes. On the control scenario, which is a static 5-fold cross-validation evaluation, the \textit{Graph set} model exhibits an higher MCC than the \textit{Base set} model over all the capture period. Whereas, on the scenario 3, which display multiple models behaviors on short-term drift, the \textit{Graph set} models are consistently underperforming compared to the \textit{Base set}. Nevertheless, on the long-term drift scenarios 1 and 2, while the initial performance at 1-2 of the \textit{Graph set} models are outperformed by the \textit{Base set}, they show better average performances over time, which raises interest over graph structural metrics resilience to concept drift. 

\subsection{Graph community metrics}
\label{eval:gcmetrics}

\paragraph{Baseline evaluation on control learning scenario}

\begin{figure*}[!htb]
\begin{tabular}{cc}
\multicolumn{2}{c}{} \\
\resizebox{0.49\linewidth}{!}{\begin{tikzpicture}
\begin{axis}[xlabel={a. Learning scenario 1: 2 first days labeled},
xlabel style={align=center, text width=6cm},
ylabel={MCC},
xmin=0,
xmax=8,
ymin=0,
ymax=1,
xtick={0,1,2,3,4,5,6,7,8},
xticklabels={1-2,2-1,2-2,3-1,3-2,4-1,4-2,5-1,5-2},
ytick={0,0.1, 0.2, 0.3, 0.4, 0.5,0.6,0.7,0.8,0.9, 1},
legend pos=south west,
legend style={fill=none},
ymajorgrids=true,
grid style=dashed]
\addplot[
,color=blue,
mark=square,
]
coordinates {
(0,0.9193545620707999)
(1,0.7710134161188049)
(2,0.5744032158538639)
(3,0.48256616288914317)
(4,0.029747834168541046)
(5,0.30596327884999736)
(6,0.0297888305162494)
(7,0.026749404951051027)
(8,0.030503136339862024)
};
\addlegendentry{Base set}

\addplot[
,color=green,
mark=diamond,
]
coordinates {
(0,0.8076949989706544)
(1,0.5276967105805128)
(2,0.6067467348529397)
(3,0.4700315733593288)
(4,0.3817149113410322)
(5,0.5407882934567357)
(6,0.35198260114302465)
(7,0.20862953405018764)
(8,0.2748118937159843)
};
\addlegendentry{Graph set}

\addplot[
,color=orange,
mark=triangle,
]
coordinates {
(0,0.9133498164246976)
(1,0.7495686971490009)
(2,0.687886523531896)
(3,0.5919382659594756)
(4,0.41650383337416225)
(5,0.5003268601489217)
(6,0.26836123526928923)
(7,0.06746215959926465)
(8,0.10427552549681911)
};%0.47774143521
\addlegendentry{GC set}

\addplot[
,color=red,
mark=cross,
]
coordinates {
(0,0.7583413255868389)
(1,0.6586732350864778)
(2,0.7382367488796446)
(3,0.6056145044960002)
(4,0.31641494482380056)
(5,0.6061461921147102)
(6,0.20277374356887137)
(7,0.06829135283656489)
(8,0.1581040013048532)

};
\addlegendentry{t-robust set}

\addplot[
color=red,
style=dashed
]
coordinates {
(0,1)
(1,0.934285468)
(2,0.947353229)
(3,0.910165951)
(4,0.811581976)
(5,0.809535853)
(6,0.732086607)
(7,0.651832479)
(8,0.602571825)

};

\addplot[
color=orange,
style=dashed
]
coordinates {
(0,1)
(1,0.910340421)
(2,0.857942559)
(3,0.805480893)
(4,0.735588287)
(5,0.704289114)
(6,0.645650793)
(7,0.574177237)
(8,0.523065124)

};

\addplot[
color=blue,
style=dashed
]
coordinates {
(0,1)
(1,0.919323212)
(2,0.821145358)
(3,0.74708319)
(4,0.604138013)
(5,0.558915388)
(6,0.483699175)
(7,0.426873759)
(8,0.383129882)

};

\end{axis}
\end{tikzpicture}}
\resizebox{0.49\linewidth}{!}{\begin{tikzpicture}
\begin{axis}[xlabel={b. Learning scenario 2: 2 first days and 1-2 are labeled},
xlabel style={align=center, text width=6cm},
ylabel={MCC},
xmin=0,
xmax=8,
ymin=0,
ymax=1,
xtick={0,1,2,3,4,5,6,7,8},
xticklabels={1-2,2-1,2-2,3-1,3-2,4-1,4-2,5-1,5-2},
ytick={0,0.1, 0.2, 0.3, 0.4, 0.5,0.6,0.7,0.8,0.9, 1},
legend pos=north east,
legend style={fill=none},
ymajorgrids=true,
grid style=dashed]
\addplot[
,color=blue,
mark=square,
]
coordinates {
(1,0.783304801120825)
(2,0.5914762532452313)
(3,0.5172663334374789)
(4,0.026158313453209892)
(5,0.3118288029422478)
(6,0.027137499093373086)
(7,0.027074207482488132)
(8,0.043142635980527345)

};
\addlegendentry{Base set}

\addplot[
,color=green,
mark=diamond,
]
coordinates {
(1,0.6440615217991053)
(2,0.6185818254436102)
(3,0.37679811518891)
(4,0.18773205303558077)
(5,0.5063009110521935)
(6,0.11027302589222153)
(7,0.056555678744612245)
(8,0.02172925215363199)
};
\addlegendentry{Graph set}

\addplot[
,color=orange,
mark=triangle,
]
coordinates {
(1,0.7865179335716289)
(2,0.486109709475969)
(3,0.30926332123770417)
(4,0.04943294816637151)
(5,0.30020372640561327)
(6,0.039480512576950795)
(7,0.023153517889032027)
(8,0.046066219662191106)
};%0.25502848612
\addlegendentry{GC set}

\addplot[
,color=red,
mark=cross,
]
coordinates {
(1,0.7255439134034775)
(2,0.35827585892369185)
(3,0.24429869706456483)
(4,0.10775145517299255)
(5,0.254095604558384)
(6,0.0323256913144209)
(7,0.0174996349043579)
(8,0.07829036480725274)

};
\addlegendentry{t-robust set}

\addplot[
color=blue,
style=dashed
]
coordinates {
(1,1)
(2,0.877551786)
(3,0.805155875)
(4,0.612215608)
(5,0.569391251)
(6,0.480266856)
(7,0.41659503)
(8,0.371405365)

};

\addplot[
color=orange,
style=dashed
]
coordinates {
(1,1)
(2,0.809026463)
(3,0.670419536)
(4,0.518527246)
(5,0.518527246)
(6,0.424861113)
(7,0.358957262)
(8,0.316044771)

};

\addplot[
color=red,
style=dashed
]
coordinates {
(1,1)
(2,0.746901567)
(3,0.610171415)
(4,0.494756381)
(5,0.465847896)
(6,0.395632203)
(7,0.342558936)
(8,0.313227288)

};

\end{axis}
\end{tikzpicture}}
\\
\resizebox{0.49\linewidth}{!}{\begin{tikzpicture}
\begin{axis}[xlabel={c. Learning scenario 3: Training on previous period labeled},
xlabel style={align=center, text width=6cm},
ylabel={MCC},
xmin=0,
xmax=8,
ymin=0,
ymax=1,
xtick={0,1,2,3,4,5,6,7,8},
xticklabels={1-2,2-1,2-2,3-1,3-2,4-1,4-2,5-1,5-2},
ytick={0,0.1, 0.2, 0.3, 0.4, 0.5,0.6,0.7,0.8,0.9, 1},
legend pos=south east,
legend style={fill=none},
ymajorgrids=true,
grid style=dashed]
\addplot[
,color=blue,
mark=square,
]
coordinates {
(0,0.9193545620707999)
(1,0.7805216341716685)
(2,0.33561964952303974)
(3,0.9855320139121546)
(4,0.04640183322624532)
(5,0.9157274670469293)
(6,0.8987953161799183)
(7,0.9330360686727948)
(8,0.7780854006362011)
};
\addlegendentry{Base set}

\addplot[
,color=green,
mark=diamond,
]
coordinates {
(0,0.8076949989706544)
(1,0.6347625068885158)
(2,0.28680433538739414)
(3,0.9507265061647522)
(4,0.09681671518668536)
(5,0.9071047683726555)
(6,0.8666370207272365)
(7,0.9037996859283661)
(8,0.7661352123915406)
};
\addlegendentry{Graph set}

\addplot[
,color=orange,
mark=triangle,
]
coordinates {
(0,0.9133498164246976)
(1,0.7808889698149932)
(2,0.3286729503438601)
(3,0.9831575339478705)
(4,0.10516899627464198)
(5,0.8517681453849729)
(6,0.5717276480258818)
(7,0.9522887705558496)
(8,0.7821051224053575)

};
\addlegendentry{GC set}

\addplot[
,color=red,
mark=cross,
]
coordinates {
(0,0.7583413255868389)
(1,0.6941142300977755)
(2,0.28766513448598413)
(3,0.972345341580207)
(4,0.0875580651793371)
(5,0.7600414959848972)
(6,0.7321603047044775)
(7,0.9051680197208307)
(8,0.7719513702971919)

};
\addlegendentry{t-robust set}

\addplot[
color=red,
style=dashed
]
coordinates {
(0,1)
(1,0.957652911)
(2,0.764880154)
(3,0.894210147)
(4,0.738460111)
(5,0.782423752)
(6,0.808574065)
(7,0.856704254)
(8,0.874620128)

};

\addplot[
color=blue,
style=dashed
]
coordinates {
(0,1)
(1,0.92449435)
(2,0.73801626)
(3,0.821507823)
(4,0.667300695)
(5,0.722093036)
(6,0.75859936)
(7,0.790634645)
(8,0.796823991)

};

\addplot[
color=orange,
style=dashed
]
coordinates {
(0,1)
(1,0.927486247)
(2,0.738275631)
(3,0.82281433)
(4,0.681280756)
(5,0.723163302)
(6,0.709278241)
(7,0.750947601)
(8,0.76265387)

};

\end{axis}
\end{tikzpicture}}
\resizebox{0.49\linewidth}{!}{\begin{tikzpicture}
\begin{axis}[xlabel={d. Control learning scenario : classification 5-folds on same period},
xlabel style={align=center, text width=8cm},
ylabel={MCC},
xmin=0,
xmax=8,
ymin=0.75,
ymax=1,
xtick={0,1,2,3,4,5,6,7,8},
xticklabels={1-2,2-1,2-2,3-1,3-2,4-1,4-2,5-1,5-2},
ytick={0.7,0.8,0.9, 1},
legend pos=south west,
legend style={fill=none},
ymajorgrids=true,
grid style=dashed]
\addplot[
,color=blue,
mark=square,
]
coordinates {
(0,0.9275454268334504)
(1,0.9185307084617481)
(2,0.9868858174386907)
(3,0.991287956812895)
(4,0.9585002386365383)
(5,0.945481938643732)
(6,0.9300389238975034)
(7,0.9408694740847506)
(8,0.8037481456373003)

};
\addlegendentry{Base set}

\addplot[
,color=green,
mark=diamond,
]
coordinates {
(0,0.9913733372304755)
(1,0.9873440268998287)
(2,0.9973351241178678)
(3,0.9982075293705721)
(4,0.9952792125832444)
(5,0.9803569821274147)
(6,0.9835602116071197)
(7,0.9776377033306787)
(8,0.9728268517241554)
};
\addlegendentry{Graph set}

\addplot[
,color=orange,
mark=triangle,
]
coordinates {
(0,0.9896040352716466)
(1,0.9876333303121341)
(2,0.9973490062320869)
(3,0.998132928152842)
(4,0.9952339229089997)
(5,0.9814553809025256)
(6,0.9839002864817628)
(7,0.9799615510144386)
(8,0.9751306779704535)

};
\addlegendentry{GC set}

\addplot[
,color=red,
mark=cross,
]
coordinates {
(0,0.9538604206211556)
(1,0.9552454335561992)
(2,0.9901511922290798)
(3,0.9940137732125729)
(4,0.9917725850463697)
(5,0.9763194228958977)
(6,0.9694809873445603)
(7,0.9748075270729059)
(8,0.9694521428619005)

};
\addlegendentry{t-robust set}

\end{axis}
\end{tikzpicture}}
\end{tabular}
\caption{MCC evolution over the different scenarios for base, graph, graph community, and t-robust sets on the XGboost model used. The x-axis denotes the capture period (week-segment notation N-M) according to scenarios with N the capture week and $M \in {\{1, 2\}}$ the week period. Dashed lines are retained expectancies}
\label{fig:mcc}
\end{figure*}
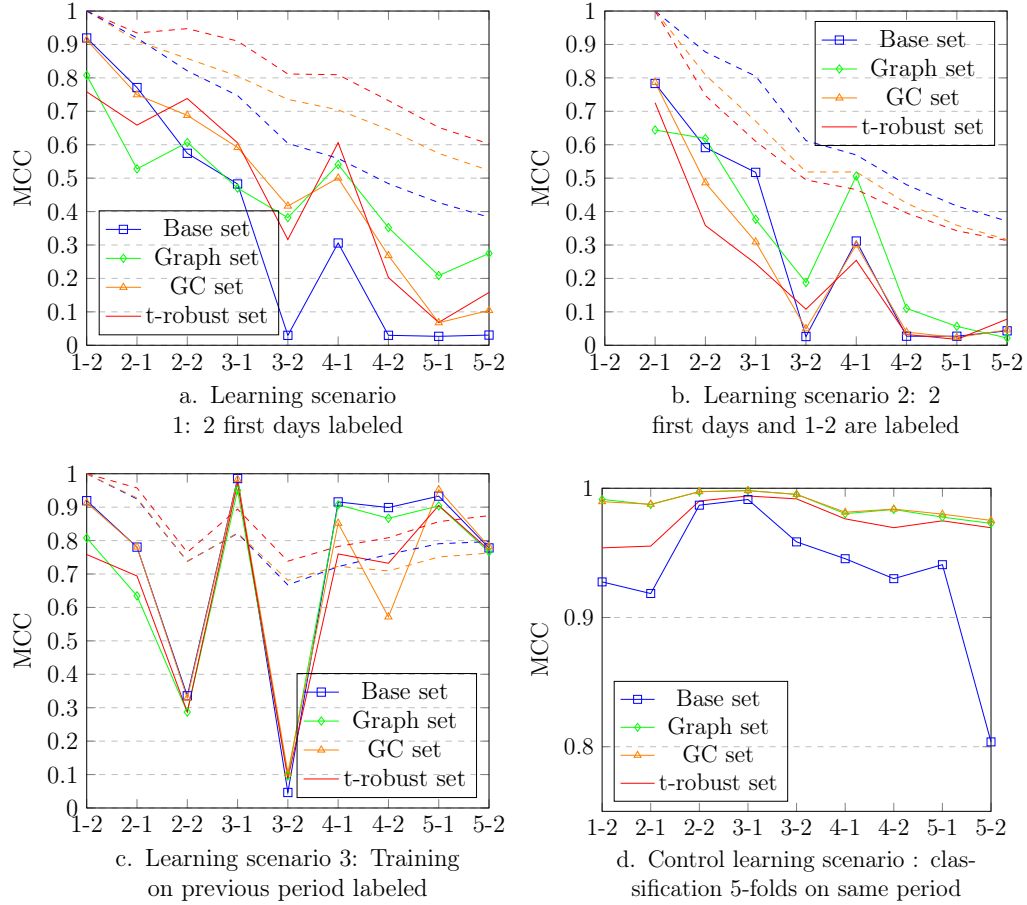

The control learning scenario of Figure~\ref{fig:mcc} evaluates the capability of the learning model in the absence of concept drift: detection is evaluated by 5-fold cross-validation applied independently to each time segment. What it measures is therefore a \textit{performance ceiling}, the detection capability attainable by each feature space under stable conditions. It answers the question which the robustness measurements cannot answer on their own, namely whether robustness has been obtained at the expense of discriminative power.
The enriched feature set, referred to as the GC set, attains the highest ceiling, with an \textbf{average MCC} of approximately $0.989$ over the dataset, against $0.9336$ for the base feature set, and $0.987$ for the Graph set. The time-robust set, or t-robust set, reaches $0.9750$. Selection on temporal robustness thus costs less than one point of MCC with respect to the full enriched set and remains more than four points above the base set: reducing the feature space does not degrade classification quality under stable conditions.

\paragraph{Selection of robust features}

\begin{table*}[!htb]
\begin{center}
\begin{tabular}{l|l|l|c|}
\cline{2-4}
\multicolumn{1}{c|}{\textit{\textbf{Features}}}                                 & \multicolumn{1}{c|}{\textbf{1-1  Coverage}} & \multicolumn{1}{c|}{\textbf{t-robustness}} & \textbf{Selected} \\ \hline
\multicolumn{1}{|l|}{{\color[HTML]{242424} \textbf{ToS}}}                       & {\color[HTML]{5F6368} 1157}                 & {\color[HTML]{242424} 0.99984}  & Yes                \\ \hline
\multicolumn{1}{|l|}{{\color[HTML]{242424} \textbf{Source Port}}}               & {\color[HTML]{5F6368} 10544}                & {\color[HTML]{242424} 0.96945}  & Yes                \\ \hline
\multicolumn{1}{|l|}{{\color[HTML]{242424} \textbf{Nb\_of\_nodes\_c\_ipport5}}} & {\color[HTML]{5F6368} 70}                   & {\color[HTML]{242424} 0.93705}  & Too low coverage  \\ \hline
\multicolumn{1}{|l|}{{\color[HTML]{242424} \textbf{Nb\_of\_edges\_c\_ipport5}}} & {\color[HTML]{5F6368} 167}                  & {\color[HTML]{242424} 0.93680}  & Too correlated    \\ \hline
\multicolumn{1}{|l|}{{\color[HTML]{242424} \textbf{delta\_Node\_ipport20}}}     & {\color[HTML]{5F6368} 102}                  & {\color[HTML]{242424} 0.92512}  & Yes                \\ \hline
\multicolumn{1}{|l|}{{\color[HTML]{242424} \textbf{Duration}}}                  & {\color[HTML]{5F6368} 458}                  & {\color[HTML]{242424} 0.90899}  & Yes                \\ \hline
\multicolumn{1}{|l|}{{\color[HTML]{242424} \textbf{delta\_expansion\_ipport5}}} & {\color[HTML]{5F6368} 0}                    & {\color[HTML]{242424} 0.89568}  & Too low coverage  \\ \hline
\multicolumn{1}{|l|}{{\color[HTML]{242424} \textbf{delta\_Node\_ipport5}}}      & {\color[HTML]{5F6368} 17017}                & {\color[HTML]{242424} 0.89492}  & Yes                \\ \hline
\multicolumn{1}{|l|}{{\color[HTML]{242424} \textbf{edges\_dist\_ipport5}}}      & {\color[HTML]{5F6368} 290}                  & {\color[HTML]{242424} 0.88725}  & Yes                \\ \hline
\multicolumn{1}{|l|}{{\color[HTML]{242424} \textbf{stability\_ipport5}}}        & {\color[HTML]{5F6368} 203}                  & {\color[HTML]{242424} 0.88429}   & Yes                \\ \hline
\end{tabular}
\end{center}
\caption{Highest \textit{t-robustness} features in UGR16 dataset, with coverage from XGBoost training on the 1-1 time interval, and rule for dropping columns: a feature is dropped if its coverage in training is below 100, or if it is too strongly correlated with a feature of higher coverage or with a feature already dropped}
\label{tab:feature}
\end{table*}

A robust dataset requires a restricted set of features. The selection, set out in Table~\ref{tab:feature}, rests on two criteria --- the \textit{t-robustness} of a feature and its coverage in the XGBoost model trained over the 1-1 time interval --- which are the operational counterparts of the two conditions stated in Section~\ref{condrift}: stability over time, and usefulness for detection.
The procedure comprises three steps: 1) \textit{pre-selection}, features whose \textit{t-robustness} exceeds 0.75 are retained; 2) \textit{coverage filtering}, among these, features whose coverage exceeds 100 are retained; 3) \textit{correlation check}, a feature is discarded when it is correlated above 90\% with another feature of higher coverage, or with a feature already discarded.
Two observations follow. Features of the original set and features derived from graph communities alike reach a \textit{t-robustness} sufficient for selection while retaining high coverage, so that the robust space is not obtained by substituting derived features for original ones, but by combining them. Among the graph community metrics, those computed on graphs whose nodes are pairs of IP addresses and ports attain higher \textit{t-robustness} than the others, which corroborates, at the level of the metrics, the granularity effect established at the level of the motifs in Section~\ref{gcm:topologies}.

\paragraph{Robustness to concept drift with graphs and robust features}

Detection performance is now compared over \textit{Scenarios 1 to 3}, between the original dataset and the dataset enriched with derived features, here graph community metrics. Figure~\ref{fig:mcc} gives the MCC of the XGBoost model for each scenario.
Scenarios 1 and 2 share the same structure and differ only in the volume of labelled data available at the outset: the first two days, that is the 1-1 interval, in Scenario 1; the whole fifth week of July, that is 1-1 and 1-2, in Scenario 2. For the base feature set, performance is almost identical in the two scenarios. For the graph community set it diverges considerably, with an average MCC difference of $0.2227$ in favour of Scenario 1 --- that is, in favour of the scenario trained on less data. This inversion contradicts the expectation that a longer training period can only help, and it is what motivates the introduction of the t-robust set; we return to it in Section~\ref{discussion}.
Learning Scenario 3, in which each model is trained on the time interval immediately preceding detection, exposes short-term drift and thereby localises the periods where the behaviour of the data changes appreciably: a sharp drop affects both models between the 3-1 and 3-2 intervals. The two models follow similar trends in this scenario, with a marked drop of the GC set model at 4-2 which the base set model does not exhibit.
The decisive observation concerns the 3-2 interval. Detection degrades severely there for every model of Scenarios 1 to 3, with two exceptions: the GC set model of Scenario 1 and the t-robust model, which retain MCC scores of $0.4165$ and $0.3164$ respectively, where the base set model collapses to $0.0297$. At the point where the environment changes most, the two feature spaces built on graph community metrics and on temporal robustness therefore retain an order of magnitude more detection capability than the base space. This is one central experimental result of this study for graph community metrics.

% Mettre triptique avec MCC rate diff et Retained expectancy diff
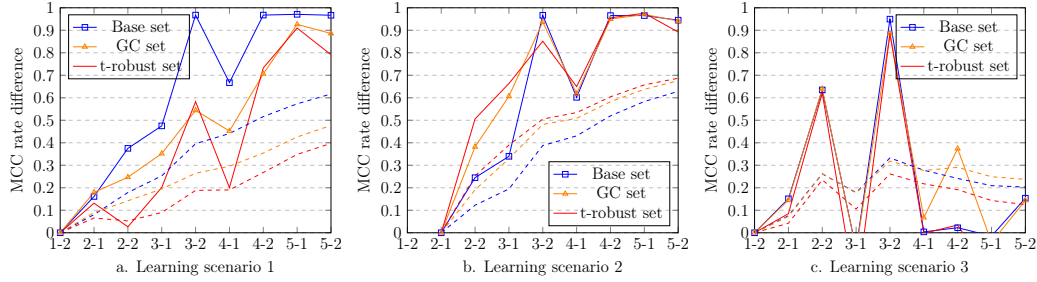
\begin{figure*}[!htb]
\begin{tabular}{ccc}
\multicolumn{3}{c}{} \\
\resizebox{\linewidth}{!}{\begin{tikzpicture}
\begin{axis}[xlabel={a. Learning scenario 1},
ylabel={MCC rate difference},
xmin=0,
xmax=8,
ymin=0,
ymax=1,
xtick={0,1,2,3,4,5,6,7,8},
xticklabels={1-2,2-1,2-2,3-1,3-2,4-1,4-2,5-1,5-2},
ytick={0,0.1, 0.2, 0.3, 0.4, 0.5,0.6,0.7,0.8,0.9, 1},
legend pos=north west,
legend style={fill=none},
ymajorgrids=true,
grid style=dashed]
\addplot[
,color=blue,
mark=square,
]
coordinates {
(0,0)
(1,0.161353576)
(2,0.375210349)
(3,0.475103314)
(4,0.967642697)
(5,0.667197737)
(6,0.967598104)	
(7,0.970904147)
(8,0.96682114)
};
\addlegendentry{Base set}
\addplot[
,color=orange,
mark=triangle,
]
coordinates {
(0,0)
(1,0.179319157)
(2,0.246853165)
(3,0.351904106)
(4,0.543982135)
(5,0.452206755)
(6,0.706179133)
(7,0.926137655)
(8,0.885831777)
};
\addlegendentry{GC set}

\addplot[
,color=red,
]
coordinates {
(0,0)
(1,0.131429064)
(2,0.02651125)
(3,0.201395883)
(4,0.582753921)
(5,0.200694764)
(6,0.732608871)
(7,0.909946418)
(8,0.791513404)
};
\addlegendentry{t-robust set}

\addplot[
,color=red,
style=dashed
]
coordinates {
(0,0)
(1,0.065714532)
(2,0.052646771)
(3,0.089834049)
(4,0.188418024)
(5,0.190464147)
(6,0.267913393)
(7,0.348167521)
(8,0.397428175)
};

\addplot[
,color=blue,
style=dashed
]
coordinates {
(0,0)
(1,0.080676788)
(2,0.178854642)
(3,0.25291681)
(4,0.395861987)
(5,0.441084612)
(6,0.516300825)
(7,0.573126241)
(8,0.616870118)
};

\addplot[
,color=orange,
style=dashed
]
coordinates {
(0,0)
(1,0.089659579)
(2,0.142057441)
(3,0.194519107)
(4,0.264411713)
(5,0.295710886)
(6,0.354349207)
(7,0.425822763)
(8,0.476934876)
};

\end{axis}
\end{tikzpicture}
\begin{tikzpicture}
\begin{axis}[xlabel={b. Learning scenario 2},
ylabel={MCC rate difference},
xmin=0,
xmax=8,
ymin=0,
ymax=1,
xtick={0,1,2,3,4,5,6,7,8},
xticklabels={1-2,2-1,2-2,3-1,3-2,4-1,4-2,5-1,5-2},
ytick={0,0.1, 0.2, 0.3, 0.4, 0.5,0.6,0.7,0.8,0.9, 1},
legend pos=south east,
legend style={fill=none},
ymajorgrids=true,
grid style=dashed]
\addplot[
,color=blue,
mark=square,
]
coordinates {
(1,0)
(2,0.244896428)
(3,0.339635947)
(4,0.966605192)
(5,0.601906177)
(6,0.965355122)
(7,0.965435923)
(8,0.944922288)
};
\addlegendentry{Base set}
\addplot[
,color=orange,
mark=triangle,
]
coordinates {
(1,0)
(2,0.381947075)
(3,0.606794317)
(4,0.937149624)
(5,0.618312929)
(6,0.949803417)
(7,0.970561996)
(8,0.941430173)
};
\addlegendentry{GC set}

\addplot[
,color=red,
]
coordinates {
(1,0)
(2,0.506196865)
(3,0.663288889)
(4,0.85148872)
(5,0.649786043)
(6,0.955446265)
(7,0.975880667)
(8,0.892094244)
};
\addlegendentry{t-robust set}

\addplot[
,color=red,
style=dashed
]
coordinates {
(1,0)
(2,0.253098433)
(3,0.389828585)
(4,0.505243619)
(5,0.534152104)
(6,0.604367797)
(7,0.657441064)
(8,0.686772712)
};

\addplot[
,color=blue,
style=dashed
]
coordinates {

(1,0)
(2,0.122448214)
(3,0.194844125)
(4,0.387784392)
(5,0.430608749)
(6,0.519733144)
(7,0.58340497)
(8,0.628594635)
};

\addplot[
,color=orange,
style=dashed
]
coordinates {
(1,0)
(2,0.190973537)
(3,0.329580464)
(4,0.481472754)
(5,0.508840789)
(6,0.58233456)
(7,0.637795623)
(8,0.675749941)
};

\end{axis}
\end{tikzpicture}
\begin{tikzpicture}
\begin{axis}[xlabel={c. Learning scenario 3},
ylabel={MCC rate difference},
xmin=0,
xmax=8,
ymin=0,
ymax=1,
xtick={0,1,2,3,4,5,6,7,8},
xticklabels={1-2,2-1,2-2,3-1,3-2,4-1,4-2,5-1,5-2},
ytick={0,0.1, 0.2, 0.3, 0.4, 0.5,0.6,0.7,0.8,0.9, 1},
legend pos=north east,
legend style={fill=none},
ymajorgrids=true,
grid style=dashed]
\addplot[
,color=blue,
mark=square,
]
coordinates {
(0,0)
(1,0.1510113)
(2,0.63493992)
(3,-0.071982513)
(4,0.949527815)
(5,0.003945262)
(6,0.022362695)
(7,-0.014881643)
(8,0.153661239)
};
\addlegendentry{Base set}
\addplot[
,color=orange,
mark=triangle,
]
coordinates {
(0,0)
(1,0.145027507)
(2,0.6401456)
(3,-0.076430428)
(4,0.884853542)
(5,0.06742397)
(6,0.37403212)
(7,-0.042633122)
(8,0.143695977)
};

\addlegendentry{GC set}

\addplot[
,color=red,
]
coordinates {
(0,0)
(1,0.084694178)
(2,0.620665359)
(3,-0.282200124)
(4,0.884540032)
(5,-0.002241959)
(6,0.034524059)
(7,-0.193615578)
(8,-0.017947123)
};
\addlegendentry{t-robust set}

\addplot[
,color=red,
style=dashed
]
coordinates {
(0,0)
(1,0.042347089)
(2,0.235119846)
(3,0.105789853)
(4,0.261539889)
(5,0.217576248)
(6,0.191425935)
(7,0.143295746)
(8,0.125379872)
};

\addplot[
,color=blue,
style=dashed
]
coordinates {
(0,0)
(1,0.07550565)
(2,0.26198374)
(3,0.178492177)
(4,0.332699305)
(5,0.277906964)
(6,0.2414006)
(7,0.209365355)
(8,0.203176009)
};

\addplot[
,color=orange,
style=dashed
]
coordinates {
(0,0)
(1,0.072513753)
(2,0.261724369)
(3,0.17718567)
(4,0.318719244)
(5,0.276836698)
(6,0.290721759)
(7,0.249052399)
(8,0.23734613)
};

\end{axis}
\end{tikzpicture}}
\end{tabular}
\caption{MCC rate difference (plain lines) and retained expectancy rate difference (dashed lines) over learning scenario 1-3 for base, graph community, and t-robust sets. Lower value indicate performance closer to initial detection.}
\label{fig:rate_difference}
\end{figure*}

\subsection{Spectral metrics}

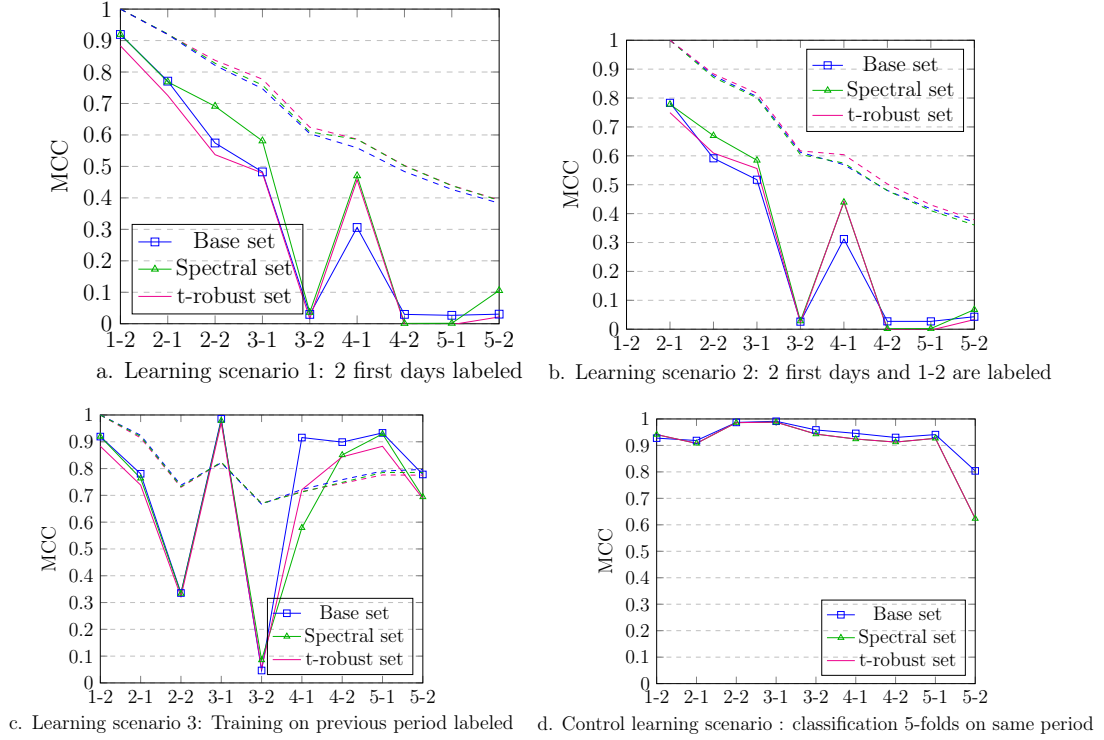
\begin{figure*}[!htb]
\begin{tabular}{cc}
\multicolumn{2}{c}{} \\
\resizebox{0.47\linewidth}{!}{\begin{tikzpicture}
\begin{axis}[xlabel={a. Learning scenario 1: 2 first days labeled},
ylabel={MCC},
xmin=0,
xmax=8,
ymin=0,
ymax=1,
xtick={0,1,2,3,4,5,6,7,8},
xticklabels={1-2,2-1,2-2,3-1,3-2,4-1,4-2,5-1,5-2},
ytick={0,0.1, 0.2, 0.3, 0.4, 0.5,0.6,0.7,0.8,0.9, 1},
legend pos=south west,
legend style={fill=none},
ymajorgrids=true,
grid style=dashed]
\addplot[
,color=blue,
mark=square,
]
coordinates {
(0,0.9193545620707999)
(1,0.7710134161188049)
(2,0.5744032158538639)
(3,0.48256616288914317)
(4,0.029747834168541046)
(5,0.30596327884999736)
(6,0.0297888305162494)
(7,0.026749404951051027)
(8,0.030503136339862024)
};
\addlegendentry{Base set}
\addplot[
,color=green!70!black,
mark=triangle,
]
coordinates {
(0,0.919586)
(1,0.768626)
(2,0.691153)
(3,0.581121)
(4,0.036200)
(5,0.470249)
(6,0.001137)
(7,0.001544)
(8,0.105583)
};%0.47774143521
\addlegendentry{Spectral set}

\addplot[
,color=magenta,
mark=cross,
]
coordinates {
(0,0.883276)
(1,0.725618)
(2,0.537299)
(3,0.478897)
(4,0.016411)
(5,0.457194)
(6,-0.002025)
(7,-0.002800)
(8,0.021631)

};
\addlegendentry{t-robust set}

\addplot[
color=magenta,
style=dashed
]
coordinates {
(0,1)
(1,0.91957715)
(2,0.837176792)
(3,0.777245801)
(4,0.624617385)
(5,0.586464593)
(6,0.503021158)
(7,0.439607286)
(8,0.393405103)

};

\addplot[
color=green!70!black,
style=dashed
]
coordinates {
(0,1)
(1,0.921117048)
(2,0.827768106)
(3,0.758456786)
(4,0.609627114)
(5,0.586229856)
(6,0.502025524)
(7,0.438765022)
(8,0.392018432)

};

\addplot[
color=blue,
style=dashed
]
coordinates {
(0,1)
(1,0.919323212)
(2,0.821145358)
(3,0.74708319)
(4,0.604138013)
(5,0.558915388)
(6,0.483699175)
(7,0.426873759)
(8,0.383129882)

};

\end{axis}
\end{tikzpicture}}
\resizebox{0.5\linewidth}{!}{\begin{tikzpicture}
\begin{axis}[xlabel={b. Learning scenario 2: 2 first days and 1-2 are labeled},
ylabel={MCC},
xmin=0,
xmax=8,
ymin=0,
ymax=1,
xtick={0,1,2,3,4,5,6,7,8},
xticklabels={1-2,2-1,2-2,3-1,3-2,4-1,4-2,5-1,5-2},
ytick={0,0.1, 0.2, 0.3, 0.4, 0.5,0.6,0.7,0.8,0.9, 1},
legend pos=north east,
legend style={fill=none},
ymajorgrids=true,
grid style=dashed]
\addplot[
,color=blue,
mark=square,
]
coordinates {
(1,0.783304801120825)
(2,0.5914762532452313)
(3,0.5172663334374789)
(4,0.026158313453209892)
(5,0.3118288029422478)
(6,0.027137499093373086)
(7,0.027074207482488132)
(8,0.043142635980527345)

};
\addlegendentry{Base set}
\addplot[
,color=green!70!black,
mark=triangle,
]
coordinates {
(1,0.778260)
(2,0.669760)
(3,0.584569)
(4,0.028327)
(5,0.439308)
(6,0.001674)
(7,0.002158)
(8,0.067781)
};%0.25502848612
\addlegendentry{Spectral set}

\addplot[
,color=magenta,
mark=cross,
]
coordinates {
(1,0.749449)
(2,0.609496)
(3,0.556057)
(4,0.021660)
(5,0.442306)
(6,-0.002260)
(7,-0.003216)
(8,0.033358)

};
\addlegendentry{t-robust set}

\addplot[
color=blue,
style=dashed
]
coordinates {
(1,1)
(2,0.877551786)
(3,0.805155875)
(4,0.612215608)
(5,0.569391251)
(6,0.480266856)
(7,0.41659503)
(8,0.371405365)

};

\addplot[
color=green!70!black,
style=dashed
]
coordinates {
(1,1)
(2,0.871477728)
(3,0.800731654)
(4,0.604627686)
(5,0.574820059)
(6,0.479486572)
(7,0.410318547)
(8,0.360585541)

};

\addplot[
color=magenta,
style=dashed
]
coordinates {
(1,1)
(2,0.88359671)
(3,0.816729095)
(4,0.617013728)
(5,0.603899453)
(6,0.502636995)
(7,0.430141363)
(8,0.378947259)

};

\end{axis}
\end{tikzpicture}}
 \\
\resizebox{0.5\linewidth}{!}{\begin{tikzpicture}
\begin{axis}[xlabel={c. Learning scenario 3: Training on previous period labeled},
ylabel={MCC},
xmin=0,
xmax=8,
ymin=0,
ymax=1,
xtick={0,1,2,3,4,5,6,7,8},
xticklabels={1-2,2-1,2-2,3-1,3-2,4-1,4-2,5-1,5-2},
ytick={0,0.1, 0.2, 0.3, 0.4, 0.5,0.6,0.7,0.8,0.9, 1},
legend pos=south east,
legend style={fill=none},
ymajorgrids=true,
grid style=dashed]
\addplot[
,color=blue,
mark=square,
]
coordinates {
(0,0.9193545620707999)
(1,0.7805216341716685)
(2,0.33561964952303974)
(3,0.9855320139121546)
(4,0.04640183322624532)
(5,0.9157274670469293)
(6,0.8987953161799183)
(7,0.9330360686727948)
(8,0.7780854006362011)
};
\addlegendentry{Base set}
\addplot[
,color=green!70!black,
mark=triangle,
]
coordinates {
(0,0.919586)
(1,0.763608)
(2,0.333546)
(3,0.978901)
(4,0.085493)
(5,0.579086)
(6,0.850738)
(7,0.929270)
(8,0.693963)
};
\addlegendentry{Spectral set}

\addplot[
,color=magenta,
mark=cross,
]
coordinates {
(0,0.883276)
(1,0.738423)
(2,0.322302)
(3,0.970805)
(4,0.053297)
(5,0.721245)
(6,0.843690)
(7,0.883101)
(8,0.688420)

};
\addlegendentry{t-robust set}

\addplot[
color=magenta,
style=dashed
]
coordinates {
(0,1)
(1,0.914713812)
(2,0.730127848)
(3,0.822166215)
(4,0.669898154)
(5,0.714280483)
(6,0.744849339)
(7,0.776140219)
(8,0.775269334)

};

\addplot[
color=blue,
style=dashed
]
coordinates {
(0,1)
(1,0.92449435)
(2,0.73801626)
(3,0.821507823)
(4,0.667300695)
(5,0.722093036)
(6,0.75859936)
(7,0.790634645)
(8,0.796823991)

};

\addplot[
color=green!70!black,
style=dashed
]
coordinates {
(0,1)
(1,0.919988092)
(2,0.732455431)
(3,0.82412106)
(4,0.669807572)
(5,0.712562156)
(6,0.748378654)
(7,0.785165741)
(8,0.785560751)

};

\end{axis}
\end{tikzpicture}}
\resizebox{0.55\linewidth}{!}{\begin{tikzpicture}
\begin{axis}[xlabel={d. Control learning scenario : classification 5-folds on same period},
ylabel={MCC},
xmin=0,
xmax=8,
ymin=0,
ymax=1,
xtick={0,1,2,3,4,5,6,7,8},
xticklabels={1-2,2-1,2-2,3-1,3-2,4-1,4-2,5-1,5-2},
ytick={0,0.1, 0.2, 0.3, 0.4, 0.5,0.6,0.7,0.8,0.9, 1},
legend pos=south east,
legend style={fill=none},
ymajorgrids=true,
grid style=dashed]
\addplot[
,color=blue,
mark=square,
]
coordinates {
(0,0.9275454268334504)
(1,0.9185307084617481)
(2,0.9868858174386907)
(3,0.991287956812895)
(4,0.9585002386365383)
(5,0.945481938643732)
(6,0.9300389238975034)
(7,0.9408694740847506)
(8,0.8037481456373003)

};
\addlegendentry{Base set}
\addplot[
,color=green!70!black,
mark=triangle,
]
coordinates {
(0,0.940971)
(1,0.907801)
(2,0.986096)
(3,0.987804)
(4,0.942813)
(5,0.924081)
(6,0.912633)
(7,0.926721)
(8,0.623554)

};
\addlegendentry{Spectral set}

\addplot[
,color=magenta,
mark=cross,
]
coordinates {
(0,0.940971)
(1,0.907801)
(2,0.986096)
(3,0.987804)
(4,0.942813)
(5,0.924081)
(6,0.912633)
(7,0.926721)
(8,0.623554)

};
\addlegendentry{t-robust set}

\end{axis}
\end{tikzpicture}}
\end{tabular}
\caption{MCC evolution over the different scenarios for base, graph spectral, and t-robust sets on the XGboost model used. The abscissas denotes the capture period, written as N-M according to scenarios with N the capture week and $M \in {\{1, 2\}}$ the week period. Dashed lines are retained expectancies}
\label{fig:mccspec}
\end{figure*}

This subsection evaluates spectral metrics under the same protocol. The evaluation is conducted on a random sample of 150 000 entries drawn from the UGR16 test dataset, a restriction imposed by the time complexity of the extraction of spectral metrics. The sample is thus a constraint of the method rather than a design choice, and its effect on the results is assessed below.
The evaluation process and the experimental setup are those of Section~\ref{eval:gcmetrics}. Detection performance is compared over \textit{Scenarios 1 to 4} between the original dataset, the dataset enriched with derived spectral metrics and the robust selection, referred to respectively as the base set, the spectral set and the t-robust set. Figure~\ref{fig:mccspec} gives the MCC of the XGBoost model in each scenario.

The t-robust set is derived by the procedure of Section~\ref{eval:gcmetrics}, and comprises the following features: \newline
$rate$, $ts2\_wiriness\_bytes$, $ts2\_wiriness\_rate$, $ts1\_wiriness\_bytes$,\newline $ts1\_wiriness\_rate$, $Packets$, $Bytes$, $Duration$, $Destination Port$, $Source Port$, $ToS$.

Some base features whose \textit{t-robustness} exceeded the threshold during the extraction of graph community metrics are not selected here. The cause lies in the size of the sample drawn from the original dataset, which is insufficient to establish their stability. The selection threshold was consequently lowered and aligned on the \textit{t-robustness} of the last feature retained for the GC set, so that the two families are selected at a comparable level of stability.

In the control learning case, the three models reach ceiling performance at the 2-2 and 3-1 intervals, as shown in Figure~\ref{fig:mccspec}.d. The base set here outperforms the spectral and t-robust sets, with an average MCC of $0.9336$. The contrast with the control scenario of the graph community evaluation, where the enriched sets exceeded the base set, isolates the sample size as the cause: the spectral and t-robust sets are computed on 150 000 entries, and their lower ceiling measures the cost of that restriction.

Scenarios 1 and 2 follow the evaluation process used for graph community metrics. The model trained on the base feature set again performs almost identically in the two scenarios, whereas the spectral set again yields a higher average MCC in Scenario 1 than in Scenario 2, as shown in Figure~\ref{fig:mccspec}. The inversion observed for graph community metrics is thus reproduced on a different family of derived features, and is therefore not specific to one metric. At the 4-1 interval, the spectral and t-robust sets outperform the base set, as Figure~\ref{fig:mccspec}.a,b shows.
The MCC of all three sets drops significantly between the 3-1 and 3-2 intervals, as shown in Figure~\ref{fig:mccspec}.c, which once more localises a period of significant change in the data. Between the 4-1 and 5-2 intervals, the MCC of the t-robust and spectral sets is slightly below that of the base set.

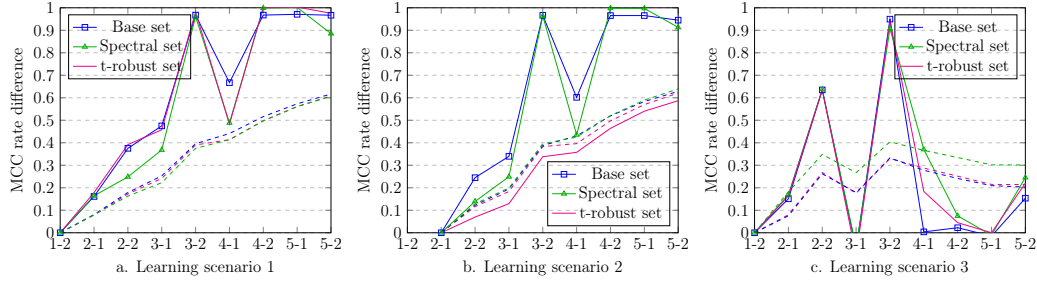
\begin{figure*}[!htb]
\begin{tabular}{ccc}
\multicolumn{3}{c}{} \\
\resizebox{\linewidth}{!}{\begin{tikzpicture}
\begin{axis}[xlabel={a. Learning scenario 1},
ylabel={MCC rate difference},
xmin=0,
xmax=8,
ymin=0,
ymax=1,
xtick={0,1,2,3,4,5,6,7,8},
xticklabels={1-2,2-1,2-2,3-1,3-2,4-1,4-2,5-1,5-2},
ytick={0,0.1, 0.2, 0.3, 0.4, 0.5,0.6,0.7,0.8,0.9, 1},
legend pos=north west,
legend style={fill=none},
ymajorgrids=true,
grid style=dashed]
\addplot[
,color=blue,
mark=square,
]
coordinates {
(0,0)
(1,0.161353576)
(2,0.375210349)
(3,0.475103314)
(4,0.967642697)
(5,0.667197737)
(6,0.967598104)	
(7,0.970904147)
(8,0.96682114)
};
\addlegendentry{Base set}
\addplot[
,color=green!70!black,
mark=triangle,
]
coordinates {
(0,0)
(1,0.164160829)
(2,0.248408523)
(3,0.368062367)
(4,0.960634459)
(5,0.488629666)
(6,0.998763574)
(7,1.001679016)
(8,0.885184202)
};
\addlegendentry{Spectral set}

\addplot[
,color=magenta,
]
coordinates {
(0,0)
(1,0.17849234)
(2,0.391697499)
(3,0.457817262)
(4,0.981420304)
(5,0.482388291)
(6,1.002292602)
(7,1.003170017)
(8,0.975510486)
};
\addlegendentry{t-robust set}

\addplot[
,color=magenta,
style=dashed
]
coordinates {
(0,0)
(1,0.078882952)
(2,0.172231894)
(3,0.241543214)
(4,0.390372886)
(5,0.413770144)
(6,0.497974476)
(7,0.561234978)
(8,0.607981568)
};

\addplot[
,color=blue,
style=dashed
]
coordinates {
(0,0)
(1,0.080676788)
(2,0.178854642)
(3,0.25291681)
(4,0.395861987)
(5,0.441084612)
(6,0.516300825)
(7,0.573126241)
(8,0.616870118)
};

\addplot[
,color=green!70!black,
style=dashed
]
coordinates {
(0,0)
(1,0.08042285)
(2,0.162823208)
(3,0.222754199)
(4,0.375382615)
(5,0.413535407)
(6,0.496978842)
(7,0.560392714)
(8,0.606594897)
};

\end{axis}
\end{tikzpicture}
\begin{tikzpicture}
\begin{axis}[xlabel={b. Learning scenario 2},
ylabel={MCC rate difference},
xmin=0,
xmax=8,
ymin=0,
ymax=1,
xtick={0,1,2,3,4,5,6,7,8},
xticklabels={1-2,2-1,2-2,3-1,3-2,4-1,4-2,5-1,5-2},
ytick={0,0.1, 0.2, 0.3, 0.4, 0.5,0.6,0.7,0.8,0.9, 1},
legend pos=south east,
legend style={fill=none},
ymajorgrids=true,
grid style=dashed]
\addplot[
,color=blue,
mark=square,
]
coordinates {
(1,0)
(2,0.244896428)
(3,0.339635947)
(4,0.966605192)
(5,0.601906177)
(6,0.965355122)
(7,0.965435923)
(8,0.944922288)
};
\addlegendentry{Base set}
\addplot[
,color=green!70!black,
mark=triangle,
]
coordinates {
(1,0)
(2,0.139413564)
(3,0.248876982)
(4,0.963602138)
(5,0.435525403)
(6,0.997849048)
(7,0.997227148)
(8,0.912906998)
};
\addlegendentry{Spectral set}

\addplot[
,color=magenta,
]
coordinates {
(1,0)
(2,0.069706782)
(3,0.129430182)
(4,0.337973171)
(5,0.357483617)
(6,0.464211189)
(7,0.540356326)
(8,0.58692516)
};
\addlegendentry{t-robust set}

\addplot[
,color=magenta,
style=dashed
]
coordinates {
(1,0)
(2,0.11640329)
(3,0.183270905)
(4,0.382986272)
(5,0.396100547)
(6,0.497363005)
(7,0.569858637)
(8,0.621052741)
};

\addplot[
,color=blue,
style=dashed
]
coordinates {

(1,0)
(2,0.122448214)
(3,0.194844125)
(4,0.387784392)
(5,0.430608749)
(6,0.519733144)
(7,0.58340497)
(8,0.628594635)
};

\addplot[
,color=green!70!black,
style=dashed
]
coordinates {
(1,0)
(2,0.128522272)
(3,0.199268346)
(4,0.395372314)
(5,0.425179941)
(6,0.520513428)
(7,0.589681453)
(8,0.639414459)
};

\end{axis}
\end{tikzpicture}
\begin{tikzpicture}
\begin{axis}[xlabel={c. Learning scenario 3},
ylabel={MCC rate difference},
xmin=0,
xmax=8,
ymin=0,
ymax=1,
xtick={0,1,2,3,4,5,6,7,8},
xticklabels={1-2,2-1,2-2,3-1,3-2,4-1,4-2,5-1,5-2},
ytick={0,0.1, 0.2, 0.3, 0.4, 0.5,0.6,0.7,0.8,0.9, 1},
legend pos=north east,
legend style={fill=none},
ymajorgrids=true,
grid style=dashed]
\addplot[
,color=blue,
mark=square,
]
coordinates {
(0,0)
(1,0.1510113)
(2,0.63493992)
(3,-0.071982513)
(4,0.949527815)
(5,0.003945262)
(6,0.022362695)
(7,-0.014881643)
(8,0.153661239)
};
\addlegendentry{Base set}
\addplot[
,color=green!70!black,
mark=triangle,
]
coordinates {
(0,0)
(1,0.169617632)
(2,0.637286779)
(3,-0.064501852)
(4,0.90703099)
(5,0.37027532)
(6,0.074868473)
(7,-0.010530826)
(8,0.2453528)
};

\addlegendentry{Spectral set}

\addplot[
,color=magenta,
]
coordinates {
(0,0)
(1,0.163995173)
(2,0.635106128)
(3,-0.099095866)
(4,0.93965985)
(5,0.183443227)
(6,0.044817249)
(7,0.000198126)
(8,0.220606017)
};
\addlegendentry{t-robust set}

\addplot[
,color=magenta,
style=dashed
]
coordinates {
(0,0)
(1,0.080011908)
(2,0.267544569)
(3,0.17587894)
(4,0.330192428)
(5,0.287437844)
(6,0.251621346)
(7,0.214834259)
(8,0.214439249)
};

\addplot[
,color=blue,
style=dashed
]
coordinates {
(0,0)
(1,0.07550565)
(2,0.26198374)
(3,0.178492177)
(4,0.332699305)
(5,0.277906964)
(6,0.2414006)
(7,0.209365355)
(8,0.203176009)
};

\addplot[
,color=green!70!black,
style=dashed
]
coordinates {
(0,0)
(1,0.181911643)
(2,0.348672808)
(3,0.267160249)
(4,0.404381664)
(5,0.366362664)
(6,0.334513273)
(7,0.301800798)
(8,0.301449541)
};

\end{axis}
\end{tikzpicture}}
\end{tabular}
\caption{MCC rate difference (plain lines) and retained expectancy rate difference (dashed lines) over learning scenario 1-3 for base, graph spectral, and t-robust sets. Lower value indicate performance closer to initial detection.}
\label{fig:rate_difference-spec}
\end{figure*}

\subsection{Properties of robust feature spaces}

The properties of the robust feature spaces are examined through the three quantities defined in Section~\ref{metodo}, which characterise performance-driven robustness to concept drift.

\textit{Worst-case performance: } for each model and each learning scenario, Figures~\ref{fig:mcc} and \ref{fig:mccspec} give the worst-case performance. In Scenario 1 (Figure~\ref{fig:mcc}.a) the base set model falls to an MCC of $0.0267$ at 5-1, whereas the t-robust feature set model reaches $0.0682$ at the same interval, which is also its worst performance in this scenario. The same pattern holds in Scenario 3 at the 3-2 period (Figure~\ref{fig:mcc}.c), where the base set model falls to an MCC of $0.0464$ while the t-robust set model holds at $0.0875$. In both cases the floor of the t-robust model lies markedly above that of the base set model, which is the property sought: what a robust feature space improves is not the peak of detection but its lower bound. Figure~\ref{fig:mccspec} exhibits no such separation, the t-robust set performing close to the base set in all 3+1 scenarios, which follows from the t-robust set selected in the spectral case containing very few base features. Sample quality and sample size therefore affect the \textit{t-robustness} of the features themselves, and not only the performance derived from them.

\textit{Retained Expectancy}: for each model of Learning Scenarios 1 to 3, we evaluate the proportion of the current performance relative to the initial performance, taken at 1-2, or at 2-1 for Scenario 2. It is represented by the dashed lines of Figures~\ref{fig:mcc} and \ref{fig:mccspec}. In Scenarios 1 and 2 the tendency is a gradual decline of retained expectancy over time, which is the expected signature of an environment subject to drift and confirms that the scenarios do expose the models to it.

\textit{Detection Rate Difference}: we assess the \textit{MCC rate difference} and the \textit{retained expectancy rate difference} with the formulas of Equations~\ref{eq:MCC_rate_difference} and \ref{eq:retained_expectancy_rate_difference}, for each model over Scenarios 1 to 3. As Figures~\ref{fig:rate_difference} and \ref{fig:rate_difference-spec} show, the MCC rate difference of the t-robust feature set is consistently lower than that of the other models, and increasingly so in the later time intervals --- the degradation of the t-robust models is not merely smaller, it grows more slowly.

Three conclusions follow from these measurements. The proposed methodology yields feature spaces which sustain detection performance in the presence of concept drift, and it does so without any update of the learning model. The enrichment of the feature space with graph metrics, whether community or spectral, improves detection, which we ascribe to their capacity to expose anomalous connectivity patterns such as those produced by scans, denials of service or endpoint-to-endpoint communications, in accordance with the motifs established in Section~\ref{gcm:topologies}. The benefit of the two families is however unequal under our experimental conditions, spectral metrics being penalised by the sample size which their computational cost imposes; this asymmetry is examined in Section~\ref{discussion}. Graph metrics are therefore a promising approach for the design of robust feature spaces, and the proposed methodology is a practical means to select them.

\section{Discussion}
\label{discussion}

This section assesses what the experiments establish. We consider in turn the instrument itself --- the quantification of drift at the level of the individual feature --- and the robustness which the feature spaces built with it exhibit.
The evaluation of current experiments and results leads us to the elicitation of novel research challenges, which will pave the way to a still better understanding of these issues to build lasting security detectors.

\subsection{Quantifying feature drift}
\label{disc:quantifying}

Feature drift does not exhaust concept drift, but it is a direct manifestation of it, and it has the property which concept drift lacks: it is observable feature by feature. \textit{State distance} and \textit{t-equivalency} measure it independently for each feature, and \textit{t-robustness} aggregates the two into a single stability score. Being bounded in $[0,1]$, that score makes features comparable and rankable according to their drift behaviour, which Table~\ref{tab:feature} illustrates. The instrument therefore provides a principled way of quantifying concept drift in communication data, that supports a verdict per feature rather than per model or per dataset.

The performance evaluations of Figures~\ref{fig:mcc} and~\ref{fig:mccspec} first confirm that UGR16 is indeed an environment subject to concept drift. Across the learning scenarios, XGBoost models which start from comparable performance diverge over time on identical data subsets. Scenario 1 in Figure~\ref{fig:mcc} shows this plainly: the base set and GC set models yield comparable MCC scores at time step 1--2 and differ significantly by time step 3--2, the GC set proving the more stable of the two. The divergence is not attributable to the models, which are identical, nor to the data, which are the same: it is attributable to the feature space, which is the only quantity that varies.

Scenario 2 extends the training data of Scenario 1, and one would therefore expect the GC set model to behave at least as well; it produces notably different results over the same evaluation window. Since the enriched feature space is sensitive to the period over which it is learnt,  it is necessary to identify features which are stable over time, and which therefore contribute to the durability of detection performance. This is the objective of the t-robust set, which is built from the \textit{t-robustness} scores of the individual features.
The two criteria under which that set is evaluated, Worst-case performance and Detection Rate Difference, are chosen accordingly to evaluate feature space stability: both bear on the stability rather than on the peak performance of the feature space.

The selection cannot rest on stability alone. Resistance to drift is necessary but not sufficient, as the limiting case shows: a feature whose value is constant over time is perfectly stable and contributes nothing to detection. Feature quality must therefore be assessed with respect to a defined detection objective, and selection must balance stability against predictive utility. We obtain the second term from the feature coverage of an XGBoost model during an initial training phase, applied to the complete feature space of the graph community-enriched or spectral metrics-enriched dataset. This dataset satisfies the initial conditions in all the learning scenarios evaluated, which makes it an admissible reference for the measurement.

\subsection{Evaluation of concept drift robustness}
\label{disc:robustness}

Characterising feature spaces through \textit{t-robustness} provides a principled way of identifying and understanding concept drift, and of evaluating the capacity of detection algorithms to learn durably. We summarise below what the experiments answer to each of the three research questions of Section~\ref{intro}.

\textit{RQ1: How can concept drift be quantified in communication data?}
The quantification proceeds along two independent approaches which corroborate each other. A performance-driven evaluation of the XGBoost model over a series of learning scenarios on the UGR16 dataset establishes the presence of drift. A distribution-driven measurement then computes the feature drift of each feature by combining \textit{state distance} and \textit{t-equivalency}, which assigns a drift value to every feature and provides a data-driven basis for evaluating temporal stability. The second route does not presuppose the first, which is what makes the resulting measurement independent of the detection model.

\textit{RQ2: How can this quantification be used to build a feature space which remains robust over time?}
The \textit{state distance} and \textit{t-equivalency} values yield a \textit{t-robustness} score for each feature. That score is combined with the information coverage of the XGBoost model in the initial training phase, which measures the relevance of the feature under the original detection conditions. Considering robustness and relevance jointly produces the t-robust feature space, whose purpose is to sustain detection performance over time. The evaluation shows that this construction raises the lower bound of performance rather than its peak, which is the property a detection system in production requires.

\textit{RQ3: Are graph community metrics and spectral metrics relevant candidates for building a long-lasting feature space for attack detection?}
For graph community metrics the answer is affirmative and specific. In internet traffic data, and network flow records in particular, features derived from communities built on combinations of IP addresses and port numbers as node identifiers obtain higher \textit{t-robustness} scores than the analogous metrics derived from full-graph structures or from graphs whose nodes are IP addresses alone. The granularity which Section~\ref{gcm:topologies} showed to be necessary for the motifs to be visible is thus also the granularity at which the derived metrics are most stable. For spectral metrics the answer is affirmative but conditional: the results are promising despite the smaller sample size on which they were obtained, while the time complexity of their computation remains a practical obstacle. Dynamic graph community metrics and spectral metrics therefore both hold potential for improving the resilience and the longevity of detection models in evolving network environments, the former under our experimental conditions, the latter subject to the computational reservation stated above.

Besides addressing the concept drift issue, such improved robustness thus provides a promising framework for detection in the presence in adversarial activity, in particular in the problem-space domain \cite{pierazzi2020intriguing,cortellazzi2025intriguing} where the attack manipulates real-world objects such as malware or malicious network packets.

\subsection{Challenges}
\label{disc:challenges}

Our experiments bring out three open issues and three further challenges, which pave the way for an extension of graph metric analyses.
%Beyond the validation of the proposed approach, the evaluations performed elicit further research questions, which sketch a road map for the continuation of this study and for the community:

\paragraph{Open issues}
Three open issues are identified, which pave the way for further investigation on \texttt{t-robust} spaces.
The results are heterogeneous across scenarios, which indicates that the parameters governing the selection are not yet sufficiently understood. The t-robust set built on graph community metrics does exhibit a better ceiling performance than the base set in the control scenario, but the spectral evaluation yielded limited results, its \textit{t-robustness} being degraded on the smaller sample which the computational cost of spectral metrics imposes. The threshold on \textit{t-robustness} is, finally, established empirically.

\paragraph{Further challenges}

\textit{Quantification of concept drift in data}:
1) How can a non-relative quantification of concept drift be obtained, going beyond the quantification of feature drift?; 2) How can the parametrisation of this approach be optimised by fine-tuning?

\textit{Extraction of robust feature spaces}:
1) How can the parametrisation of the learning approach be included in concept drift mitigation?; 2) How can additional derived features be incorporated into the feature space while maintaining robustness to concept drift?

\textit{Identification of relevant derived features as candidates for robust feature spaces}: How can the methodology be adapted to suit unsupervised approaches?

\section{Conclusions and Perspectives}
\label{conc}

We have proposed t-robust spaces, which restate the learning problem through a deliberate selection of features, combining base features with implicit features derived from the original traffic traces.
\texttt{t-robustness} is a \emph{computable, bounded, per-feature stability score} that can be evaluated without access to labels or to a downstream model.
It goes beyond generic marginal two-sample tests and their single yes/no dependence statistics, and yields a bounded score in $[0,1]$ that (i) is rankable across features, (ii) is cheap to compute at scale, and (iii) explicitly combines a \emph{local} and a \emph{cumulative} notion of stability.
By using drift-robustness as a feature-selection criterion rather than as a detection/alarm signal, the proposed approach supports a concrete domain instantiation on NetFlow and cybersecurity use cases with engineered, structurally-motivated features from graph community and spectral metrics, rather than generic synthetic 2-D benchmarks.
%The purpose of the selection is to shape data which support predictions that are more robust through time and less fragile to concept drift. Where the approaches of the literature seek to improve the learning models themselves, we hold that the durability of a detection model depends primarily on the stability of the features it is given, and not on the model. The adversarial problem of machine learning points in the same direction: the robustness of learning rests on the underlying data more than on the learning model~\cite{barreno2006can}.

The proposal has been applied to access network data affected by concept drift, on the example of UGR16. We produce datasets enriched with graph community metrics and with spectral metrics, then extract a t-robust space from the resulting feature set by means of the \textit{t-robustness} value, which expresses the stability of a feature in the dataset. The feature engineering approach which supports these spaces proceeds in four steps: extraction of the derived features which complement the base features of the target dataset; extraction of the feature states; computation of \textit{t-robustness} from the intermediate metrics \textit{state distance} and \textit{t-equivalency}; and finally detection of the target property, here cyberattacks against access network traffic. The validation compares, over the several learning scenarios, the feature spaces associated with an XGBoost model: the base set, the graph community set, the spectral set, and the corresponding t-robust sets.
Two results follow from this comparison with a baseline detection pipeline. The robustness of detection performance over time is strongly improved; and dynamic graph metrics contribute to that stability under evolving network conditions, by comparison with base NetFlow features. Quantitatively, the models built on the t-robust set are on average more stable through time. In learning scenario 1 in particular, the \textit{retained expectancy} at the last test interval remains at \textbf{0.6025} on the UGR16 dataset, against \textbf{0.5230} for the GC set and \textbf{0.3831} for the base set. The property matters beyond the figures: a detection system whose performance is predictable over time is a system which can be trusted between two retrainings.

These observations set the agenda of our future work. The criteria governing the selection parameters and the distances between features require finer tuning, which presupposes more controlled experimental environments in which the relations between concept drift, feature drift and detection model can be characterised more strictly. The influence of sampling, which our experiments show to be considerable, calls for an investigation of its own --- and suggests a reversal worth exploring, in which \textit{t-robustness} would serve as a control metric for the quality of a sample rather than being subject to it.

\section*{Aknowledgements}

The research work in this paper has been founded by French DGA RAPID under \textit{Damiage} project grant, as well as by French Région Grand-Est in \textit{XDGMed} Contrat doctoral.

%\section*{Declaration of competing interest}

%The authors declare that they have no known competing financial
%interests or personal relationships that could have appeared to influence the work reported in this paper.

%\section*{Credit authorship contribution statement}

%Julien Michel: Writing – original draft, Visualization, Validation, Software, Methodology, Investigation, Data curation, Conceptualization.

%Abdul Qadir Khan: Writing – review, editing, Conceptualization, Validation.

%Majed Jaber: Methodology, Investigation, Data curation, Conceptualization on spectral analysis topics.

%Pierre Parrend: Writing – original draft, Validation, Supervision, Project administration, Methodology, Conceptualization.

%\section*{Data availability}

%Experiments use UGR16 dataset \cite{macia2018ugr,UGR16}.

\bibliographystyle{elsarticle-num} 
\bibliography{mich.2026.conceptdrift}

\end{document}